\documentclass[aps,pre,preprint,showpacs,superscriptaddress]{revtex4}

\usepackage{indentfirst}
\usepackage{graphicx}
\usepackage{subfigure}
\usepackage{amsmath}

\begin{document}

\title{Conjugate field and fluctuation-dissipation relation for the 
dynamic phase transition in the two-dimensional kinetic Ising model}

\author{D.T. Robb}
\email[Corresponding author: ]{drobb@clarkson.edu}
\affiliation{School of Computational Science, Florida State University, Tallahassee, Florida 32306, USA}
\affiliation{Department of Physics, Clarkson University, Potsdam, New York 13699, USA}
\author{P.A. Rikvold}
\affiliation{School of Computational Science, Florida State University, Tallahassee, Florida 32306, USA}
\affiliation{Center for Materials Research and Technology and Department
of Physics, Florida State University, Tallahassee, Florida 32306-4350, USA}
\affiliation{National High Magnetic Field Laboratory, Tallahassee, Florida 32310, USA}
\author{A. Berger}
\affiliation{San Jose Research Center, Hitachi Global Storage Technologies, San Jose, California 95120, USA}
\author{M.A. Novotny}
\affiliation{Department of Physics and Astronomy and HPC$^2$ Center for Computational Sciences, Mississippi State University, Mississippi State, Mississippi 39762, USA}

\date{\today}

\begin{abstract}

The two-dimensional kinetic Ising model, when exposed to an oscillating applied
magnetic field, has been shown to exhibit a nonequilibrium,
second-order dynamic phase transition (DPT), 
whose order parameter $Q$ is the period-averaged magnetization. It has been
established that this DPT falls in the same universality class as the
equilibrium phase transition in the 
two-dimensional Ising model in zero applied field. 
Here we study for the first time the scaling of the dynamic order parameter 
with respect to a nonzero, period-averaged, magnetic `bias'
field, $H_b$, 
for a DPT produced by a square-wave applied field. We find evidence that 
the scaling exponent,  $\delta_{\mathrm{d}}$, of $H_b$ at the 
critical period of the DPT is equal to the exponent for the critical isotherm, 
$\delta _{\mathrm{e}}$, in the equilibrium Ising model. This implies that $H_b$ is a
significant component of the field conjugate to $Q$.  A finite-size scaling
analysis of the dynamic order parameter above the critical period provides 
further support for this result. We also demonstrate numerically that, for a range
of periods and values of $H_b$ in the critical region, a 
fluctuation-dissipation relation (FDR), with an effective temperature
$T_{\mathrm{eff}}\left(T, P, H_0\right)$ depending on the 
period, and possibly the temperature and field amplitude, holds for the variables 
$Q$ and $H_b$. This FDR justifies the use of the scaled variance of $Q$ as 
a proxy for the nonequilibrium susceptibility, 
$\partial \langle Q \rangle / \partial H_b$, in the critical region.

\end{abstract}


\maketitle

\section{Introduction}
\label{s:intro}
The dynamic phase transition (DPT) in a ferromagnetic system below its critical temperature was first observed in numerical solutions of 
a mean-field model exposed to an oscillating magnetic field \cite{kn:tome90, kn:mendes91}. It was then studied
further, both in mean-field models \cite{kn:zimmer93,kn:acharyya95, 
kn:acharyya97b, kn:acharyya98} and in kinetic Monte Carlo (KMC)
simulations \cite{kn:lo90, kn:acharyya97a,kn:acharyya97b,
kn:acharyya98,kn:sides98,kn:sides99}. A review
of this early work can be found in Ref.~\cite{kn:chakrabarti99}. 
More recently, the study of the DPT has expanded to include varying 
(and often more physical) 
model geometries. These include mean-field studies of domain-wall 
motion in an anisotropic XY model in one dimension 
\cite{kn:yasui03,kn:fujiwara04,kn:fujiwara06}, KMC simulations of
a three-dimensional Ising system \cite{kn:shao04}, and KMC simulations of a 
uniaxially anisotropic Heisenberg system in an off-axial field 
\cite{kn:acharyya03}, an elliptically polarized applied field 
\cite{kn:acharyya04}, and with
the effect of a thin-film surface energy \cite{kn:jang01,kn:jang03a,kn:jang03b}.
The phenomenon has also been observed in simulations of CO oxidation under
oscillating CO pressure \cite{kn:machado05a,BUEN06}. Further simulation 
studies of the DPT in the two-dimensional kinetic Ising model have
appeared \cite{kn:korniss01, kn:korniss02, kn:chatterjee03, kn:chatterjee04, 
kn:acharyya99}, as well as analytical studies of the DPT \cite{kn:fujisaka01, 
kn:tutu04,kn:meilikhov04,kn:dutta04}.

Here, we concentrate on the DPT in the two-dimensional kinetic Ising model. 
It was observed in simulations of this model that there exists a 
singularity at a critical period of the 
applied oscillating field \cite{kn:sides98,kn:sides99}, and that the critical exponents 
$\beta $ and $\gamma $ (and, with less accuracy, $\nu$) are consistent with 
the universality class of the equilibrium two-dimensional Ising 
transition in zero field \cite{kn:korniss01} 
$\left( \beta = 1/8, \gamma = 7/4, \mathrm{and}~ \nu = 1 \right)$. 
In those studies, the techniques of
finite-size scaling were extended to the study of the dynamic 
order parameter ($Q$, defined
in Sec.~\ref{s:model}) in the non-equilibrium steady state. 
This provided evidence for a diverging correlation length at a critical value of the period. In particular, because the field
conjugate to $Q$ and a fluctuation-dissipation relation were not known, a susceptibility could not be measured directly, and the scaled variance 
$X_L^Q = L^2 \left(\langle Q^2\rangle - \langle Q \rangle ^2 \right)$, where $L$ is the linear system size,
was used as a proxy. 
An analytical argument, based on the correspondence of the two-dimensional kinetic Ising model and the continuous, two-dimensional
Ginzburg-Landau model at the equilibrium critical point, provided an effective Hamiltonian for the non-equilibrium system
and confirmed that the DPT is in the Ising universality class \cite{kn:fujisaka01}.
These findings are consistent with earlier symmetry arguments that any continous phase transition in a stochastic cellular automaton that preserves the Ising up-down symmetry should be in the
equilibrium Ising universality class \cite{kn:grinstein85,kn:bassler94}.

Recently, experiments were performed on a 
$\left[ \mathrm{Co(0.4 nm)}/\mathrm{Pt(0.7 nm)} \right]_3$ multilayer film
with strong uniaxial anisotropy \cite{kn:robb06}, 
whose equilibrium behavior is known to be Ising-like 
\cite{kn:mills88, kn:back95}. The film
was exposed to an oscillating (sawtooth) applied field with
varying period, in the presence of constant `bias' magnetic
fields $H_b$ of varying strength and sign. (The bias field is defined 
explicitly in Sec.~\ref{s:model}.) The behaviors of the dynamic order
parameter and its variance, as functions of the applied field period and the
bias field, provided strong evidence for the presence of the DPT in this
experimental system. The observed behavior of the order parameter with respect to the
bias field supported previous conjectures that the conjugate field could
include the period-averaged magnetic field as an important component,
and stimulated the numerical investigations in this paper. 

This paper is organized as follows. In Sec.~\ref{s:model}, we describe the 
two-dimensional kinetic Ising model and our computational methods. 
In Sec.~\ref{s:delta.scaling},
we verify directly the scaling of the dynamic order parameter
with respect to the period-averaged magnetic field at the critical period, 
with scaling exponent $\delta _{\mathrm{d}} \approx \delta _{\mathrm{e}} = 15$,
in agreement with the equilibrium Ising transition. 
In Secs.~\ref{s:finite.size} and \ref{s:first.compare}, we derive the
expected asymptotic scaling functions in a finite-size scaling analysis of 
the dynamic phase transition with non-zero period-averaged bias field, and then
compare the expected scaling of the dynamic order parameter to our numerical
results. In Sec.~\ref{s:fluct.relation}, we present numerical data assessing
the applicability of a fluctuation-dissipation relation (FDR) to
this far-from-equilibrium system. We then in Sec.~\ref{s:second.compare}
compare the expected scaling of the susceptibility (of the dynamic
order parameter) to our numerical data,
using the results of Sec.~\ref{s:fluct.relation} to reconcile our findings with
previous results on the scaling of the fluctuations of the dynamic order
parameter. Finally, we present a summary of our results in Sec.~\ref{s:conclusion}.

\section{Computational model \label{s:model}}

In order to facilitate comparison with previous results, we employ the same model 
and computational method as in Ref.~\cite{kn:korniss01}.  Specifically, 
we perform kinetic Monte Carlo (KMC) simulations of a two-dimensional periodic square 
lattice of Ising spins $S_i$, which can take only the values $S_i = \pm 1.$ The Hamiltonian 
of the model is
\begin{equation}
\mathcal{H} = -J\sum_{\langle i,j \rangle}S_i S_j - H(t) \sum _i S_i , 
\end{equation}
where $J>0$ is the ferromagnetic exchange interaction, $\sum_{\langle i,j \rangle}$ runs 
over all nearest-neighbor pairs, $\sum_{i}$ runs over
all $L^2$ lattice sites, and $H(t)$ is an oscillating, spatially uniform applied magnetic 
field. The form of $H(t)$ is here taken
as a square wave with amplitude $H_0 = 0.3J$ and period $P$, measured in Monte Carlo
steps per spin (MCSS). The square-wave form not only allows for more efficient KMC 
simulation, but also reduces the critical 
period and the finite-size effects for the DPT \cite{kn:korniss01}. 
Other symmetric field shapes, such as
sinusoidal \cite{kn:sides98,kn:sides99} and sawtooth \cite{kn:robb06}, yield essentially the same results,
but with a larger critical period and with stronger finite-size effects.
The Glauber single-spin-flip MC algorithm with updates at randomly chosen sites is used, 
in which each attempted spin flip is accepted with probability 
\begin{equation}
W\left(S_i \rightarrow -S_i\right) = \frac{1}{1+\exp \left( \Delta E / T \right)},
\end{equation}
where $\Delta E$ is the energy change that would result from acceptance of the spin flip,
and $T$ is the absolute temperature in energy units (i.e., with Boltzmann's constant set 
to unity). All simulations were performed at $T=0.8T_c$, where $T_c = 2.269J$ is
the equilibrium critical temperature of the square-lattice Ising ferromagnet in
zero applied field \cite{ONSA44}.

The system responds to the oscillating field via the time-dependent magnetization per site,
\begin{equation}
m(t) = \frac{1}{L^2} \sum_{i=1}^{L^2} S_i (t).
\end{equation}
The dynamic order parameter is defined as the average of $m(t)$
over a given field cycle $i$ \cite{kn:tome90} :
\begin{equation}
Q_i = \frac{1}{P} \int_{(i-1)P}^{iP} m(t)dt.
\end{equation} 
We define the bias field, so named because it measures the shift (or `bias') of the 
periodic field toward either negative or positive field values, as the
period-averaged magnetic field,
\begin{equation}
H_b = \frac{1}{P} \int_{0}^{P} H(t)dt.
\label{bias.def}
\end{equation} 
This definition applies generally to any periodic magnetic field $H(t)$. In this paper, the
applied field consists of a square wave with period $P$ superposed with a constant
magnetic field. Applying (\ref{bias.def}), since the period-average of the 
square-wave field is zero, the bias field $H_b$ in this case 
is simply equal to the superposed constant magnetic field. 

\section{Scaling with respect to the bias field \label{s:delta.scaling}}

In the two-dimensional equilibrium Ising model, the critical
isotherm is given (in the thermodynamic limit, i.e., as
$L \rightarrow \infty)$ as
\begin{equation}
m \left( T=T_c, H \rightarrow 0\right) \propto H^{1/\delta_\mathrm{e}},
\label{equ.powerlaw}
\end{equation}
where the critical exponent $\delta_{\mathrm{e}} = 15$ \cite{kn:landau00}.
For finite systems, this relationship breaks down when the infinite-system
correlation length, $\xi _{\infty} \left(T=T_c,H \right)$, which diverges as
$H \rightarrow 0$, becomes comparable to the linear
system size $L$. The relationship also naturally breaks down at larger fields
away from the critical region. Therefore, a plot of $m$ vs $H$ for a given system size $L$ 
will follow the power law
(\ref{equ.powerlaw}) for a range of $H$ near the critical value $H = 0$, with this range
extending to smaller $H$ as $L$ is increased \cite{kn:binder84}.  

\begin{figure}[ht]
\begin{center}
\includegraphics[width=4.5in,angle=270]{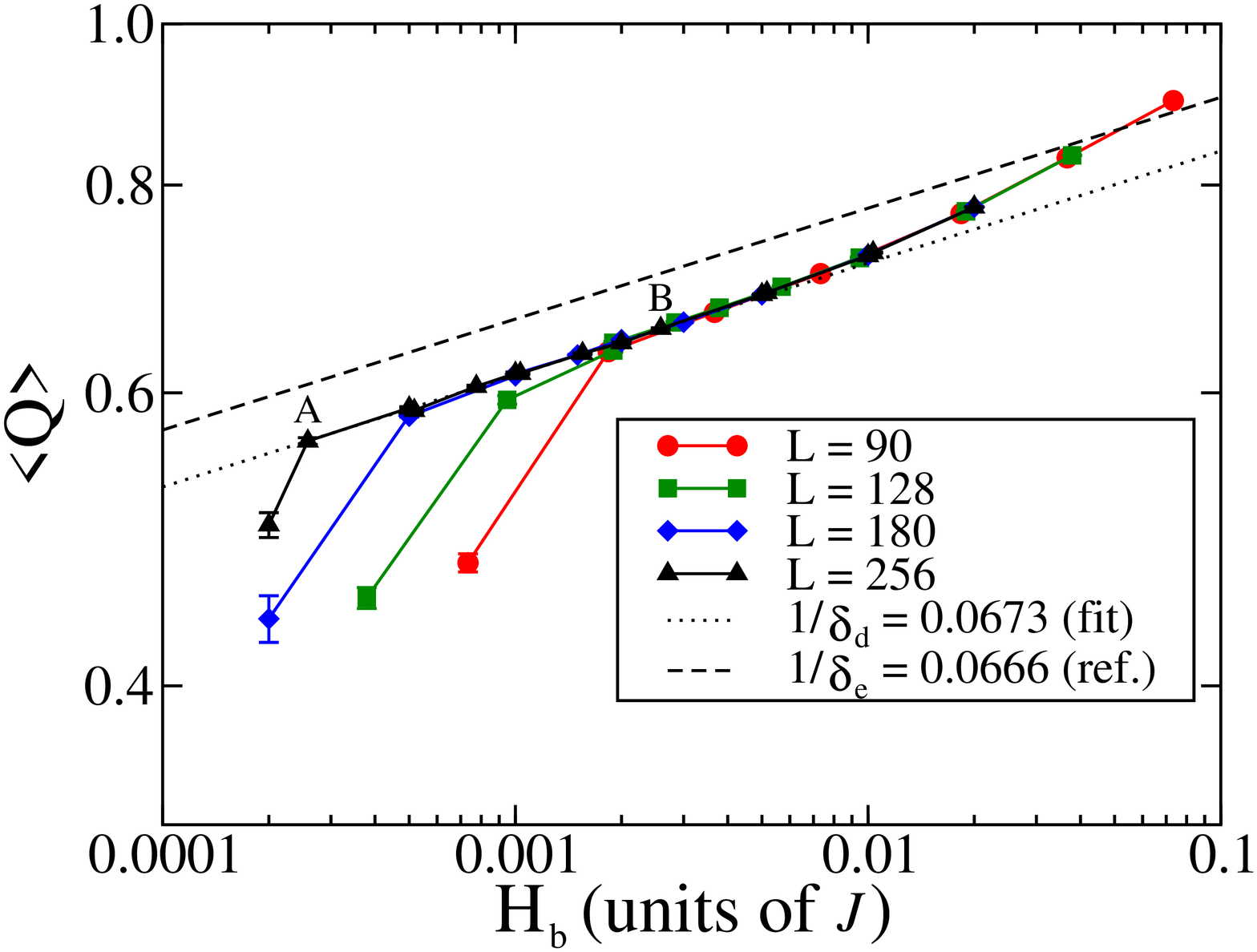} 
\end{center}
\caption
{ 
(Color online.) Log-log plot of the dynamic order parameter, $\langle Q \rangle$, vs bias field,
$H_b$, at $P = P_c$ for $L=90, 128, 180,$ and 256. A least-squares fit to power-law scaling of the $L=256$ data, in the range between the labels A and B above, produced a statistically significant fit with scaling exponent 
$1/\delta_{\mathrm{d}} = 0.0673 \pm 0.0008$ (corresponding to 
$\delta_{\mathrm{d}} = 14.85 \pm 0.18$). The dotted line corresponds to the 
scaling exponent $1/\delta_{\mathrm{d}} = 0.0673$. A reference line
representing scaling with the equilibrium Ising exponent, 
$\delta_{\mathrm{e}} = 15$ 
($1 / \delta_{\mathrm{e}} = 0.0666$), is shown as the dashed line.
}
\label{delta.scaling} 
\end{figure}

We can determine directly whether the non-equilibrium system exhibits a similar relationship,
\begin{equation}
\langle Q \rangle \left(P = P_c, H_b \rightarrow 0\right) \propto H_b^{1/\delta_{\mathrm{d}}}
\label{q.scaling}
\end{equation}
in an analogous way. In Fig.~\ref{delta.scaling}, we plot $\langle Q \rangle$
vs $H_b$ at the critical value of the period, $P = P_c$. In previous work,
the reversal time for the magnetization, following instantaneous reversal of the uniform
magnetic field $H$ at $\left(H=0.3J, T=0.8T_c\right)$, was found as 
$\tau = 74.5977$~MCSS \cite{kn:sides98,kn:sides99}. The critical scaled half-period for the
square waveform was determined to be $\Theta _c = P_c / \left( 2 \tau \right) =
0.918 \pm 0.005$ \cite{kn:korniss01}. 
This yields $P_c = 136.96 \pm 0.75$~MCSS, and in our simulations and analysis
in this paper we use  $P_c = 136.96$~MCSS.

A power-law dependence is indeed seen to hold in Fig.~\ref{delta.scaling}, 
within a range which extends to lower values of $H_b$ as $L$ is increased.  
We fit the $L=256$ data between the points labeled
A and B in Fig.~\ref{delta.scaling}, finding a statistically significant fit with 
power-law exponent $\delta_{\mathrm{d}} = 14.85 \pm 0.18$.
As including points with $H_b \ge 0.01J$ was found to greatly
reduce the statistical significance of the fit, the value $H_b = 0.01J$ serves
as a boundary of the scaling region at $P = P_c$. This result is consistent
with an exponent $\delta_{\mathrm{d}} = \delta_{\mathrm{e}} = 15$, suggesting
that the bias field $H_b$, for these parameters and the square waveform, is
the dominant component of a conjugate field which exhibits the same scaling
exponent in the DPT as does the applied magnetic field in the equilibrium
Ising transition.

\section{Finite-size scaling analysis with bias field}
\label{s:finite.size}

To provide more complete evidence that $H_b$ is the dominant component of
the field conjugate to $\langle Q \rangle$, in the next several sections
we will demonstrate data collapse onto a two-parameter finite-size scaling
function for the system-size dependent quantity $\langle Q \rangle_L$ at points 
$\left(P \geq P_c, H_b > 0\right)$,
for lattice sizes $L$ = 90, 128, 180, and (in several cases) $L=256$, 
using the critical exponents for the equilibrium Ising system. In this section,
we briefly review the theory of finite-size scaling as it applies to this system. We then determine the
expected asymptotic forms of the scaling functions, which are compared in
later sections of the paper to our computational data.

The theory of finite-size
scaling \cite{kn:privman84,kn:privman90} states that near a continuous
phase transition, the singular part of the free-energy density 
for a $d$-dimensional system of linear size $L$ can be written as 
\begin{equation}
f_L 
\approx L^{-d} Y_{\pm} \left(|\epsilon| L^{1/\nu}, H L^{\beta \delta / \nu}\right),
\label{eq:fscale}
\end{equation}
where $\epsilon = (T-T_c)/T_c$, $H$ (in units of $k_B T$) is the
field conjugate to the order parameter, $\nu$ is the critical  
exponent for the correlation length, 
$\beta$ is the exponent for the order parameter, $\delta$ is the
exponent for the critical isotherm, and $Y_{\pm}$ are scaling functions
above (+) and below ($-$) the critical point. 
This yields for the order parameter at finite $L$ 
\begin{equation}
m_L = \frac{\partial f}{\partial H} = 
L^{- \beta / \nu} \mathcal{F}_{0\pm}\left(|\epsilon| L^{1/\nu}, 
H L^{\beta \delta / \nu}\right),
\label{eq:mscale}
\end{equation}
where the exponent for $L$ in the prefactor is obtained by using the
hyperscaling relation $d \nu = 2 - \alpha$
and the exponent equality $\alpha = 2 - \beta (\delta+1)$. 
Further differentiation yields the susceptibility, 
\begin{equation}
\chi_L = \frac{\partial m_L}{\partial H} = 
L^{\gamma / \nu} \mathcal{G}_{0\pm}\left(|\epsilon| L^{1/\nu}, 
H L^{\beta \delta / \nu}\right),
\label{eq:chiscale}
\end{equation}
where the exponent for $L$ in the prefactor is obtained by using the exponent
equality $\gamma = \beta(\delta -1)$. 

It has previously been shown analytically that the DPT for a sinusoidal
applied field, which is symmetric under $H(t) \rightarrow -H(t+P/2)$ 
and so which can safely be assumed to have
$H_c=0$, has an effective Ginzburg-Landau free-energy density in the same
universality class as the equilibrium Ising model \cite{kn:fujisaka01}.
It therefore appears reasonable to write corresponding scaling
functions for the dynamic order parameter $\langle Q \rangle$ and
its associated susceptibility $\hat{\chi}$, 
\begin{equation}
\langle Q \rangle_L = 
L^{- \beta / \nu} \mathcal{F}_{\pm}\left(|\theta| L^{1/\nu}, 
(H_c/J) L^{\beta \delta / \nu}\right) \;,
\label{eq:Qscale}
\end{equation}
and 
\begin{equation}
\hat{\chi}_L = 
L^{\gamma / \nu} \mathcal{G}_{\pm}\left(|\theta| L^{1/\nu}, 
(H_c/J) L^{\beta \delta / \nu}\right) \;,
\label{eq:chihatscale}
\end{equation}
where $\theta = (P - P_c)/P_c$, and $H_c$ is the (as yet unknown) field 
conjugate to $\langle Q \rangle$. In this paper we express $H_c$ (and $H_b$) in
units of the exchange constant, $J$, so that the second scaling parameter in Eqs.~\ref{eq:Qscale} and \ref{eq:chihatscale} is dimensionless.
The specific form, $H_c/J$, with which $H_c$ is assumed 
to enter the second scaling parameter needs more theoretical 
investigation, and could conceivably change as the theory 
of the DPT is further developed. However, this should not affect 
our conclusions  \cite{scaling.param.endnote}. 
Computational results for sinusoidal and square-wave fields, which both are
symmetric under $H(t) \rightarrow -H(t+P/2)$ and so presumably have $H_c = 0$, 
have previously confirmed the scaling behavior with respect to $\theta$. The
exponent values were determined as 
$\gamma / \nu = 1.74 \pm 0.05$, $\beta / \nu = 0.126 \pm 0.005$,
and $\nu = 0.95 \pm 0.15$
\cite{kn:korniss01}, consistent with the exact values for the
two-dimensional equilibrium Ising model, 
$\gamma = 7/4 = 1.75$, $\beta = 1/8 = 0.125$, and $\nu = 1$. 

We now determine the expected asymptotic forms of the scaling functions
$\mathcal{F}_+(y_1,y_2)$ and $\mathcal{G}_+(y_1,y_2)$, 
where we emphasize that the $+$ subscript indicates that the
scaling functions refer to the
range $P \geq P_c$, and where the scaling parameters are
$y_1 \equiv \theta L^{1/ \nu}$ and 
$y_2 \equiv (H_c/J) L^{\beta \delta / \nu}$. 
\\
\underbar{$y_1 \gg y_2$}. 
We expect 
$\hat{\chi}_L \sim \theta^{-\gamma}  
= L^{\gamma / \nu} y_1^{- \gamma}$ (independent of $y_2$)
and 
$\langle Q \rangle_L = \hat{\chi}_L H_c \sim \theta^{-\gamma} H_c
\sim L^{- \beta / \nu} y_1^{- \gamma} y_2$, 
where $\gamma=\beta
(\delta -1)$ was used to obtain the exponent for $L$ in 
$\langle Q \rangle_L$. 
\\
\underbar{$y_1 \ll y_2$}.
We expect that $\langle Q \rangle_L \sim H_c^{1/\delta}  
\sim L^{- \beta / \nu} y_2^{1/\delta}$ 
and 
$\hat{\chi}_L = \partial \langle Q \rangle_L / \partial H_c 
\sim L^{\gamma / \nu} y_2^{(1-\delta)/\delta}$
(both independent of $y_1$), 
where $\gamma=\beta (\delta -1)$ 
was used to obtain the exponent for $L$ in $\hat{\chi}_L$. 
\\
Thus, the asymptotic forms of the scaling functions are expected to be
\begin{equation}
\mathcal{F}_+(y_1,y_2) \equiv L^{\beta / \nu} \langle Q \rangle_L \sim
\left\{
\begin{array}{lll}
y_1^{-\gamma} y_2    & \mbox{for} & y_1 \gg y_2 \\
y_2^{1/\delta} & \mbox{for} & y_1 \ll y_2
\end{array}
\right.
\label{eq:Fscal}
\end{equation}
and 
\begin{equation}
\mathcal{G}_+(y_1,y_2) \equiv L^{- \gamma / \nu} \hat{\chi}_L \sim
\left\{
\begin{array}{lll}
y_1^{-\gamma}             & \mbox{for} & y_1 \gg y_2 \\
y_2^{(1-\delta)/\delta} & \mbox{for} & y_1 \ll y_2
\end{array}
\right.
\;.
\label{eq:Gscal}
\end{equation}

\section{Comparison of first scaling function to computational results}
\label{s:first.compare}

\begin{figure}[ht]
\begin{center}
\includegraphics[width=3.5in,angle=270]{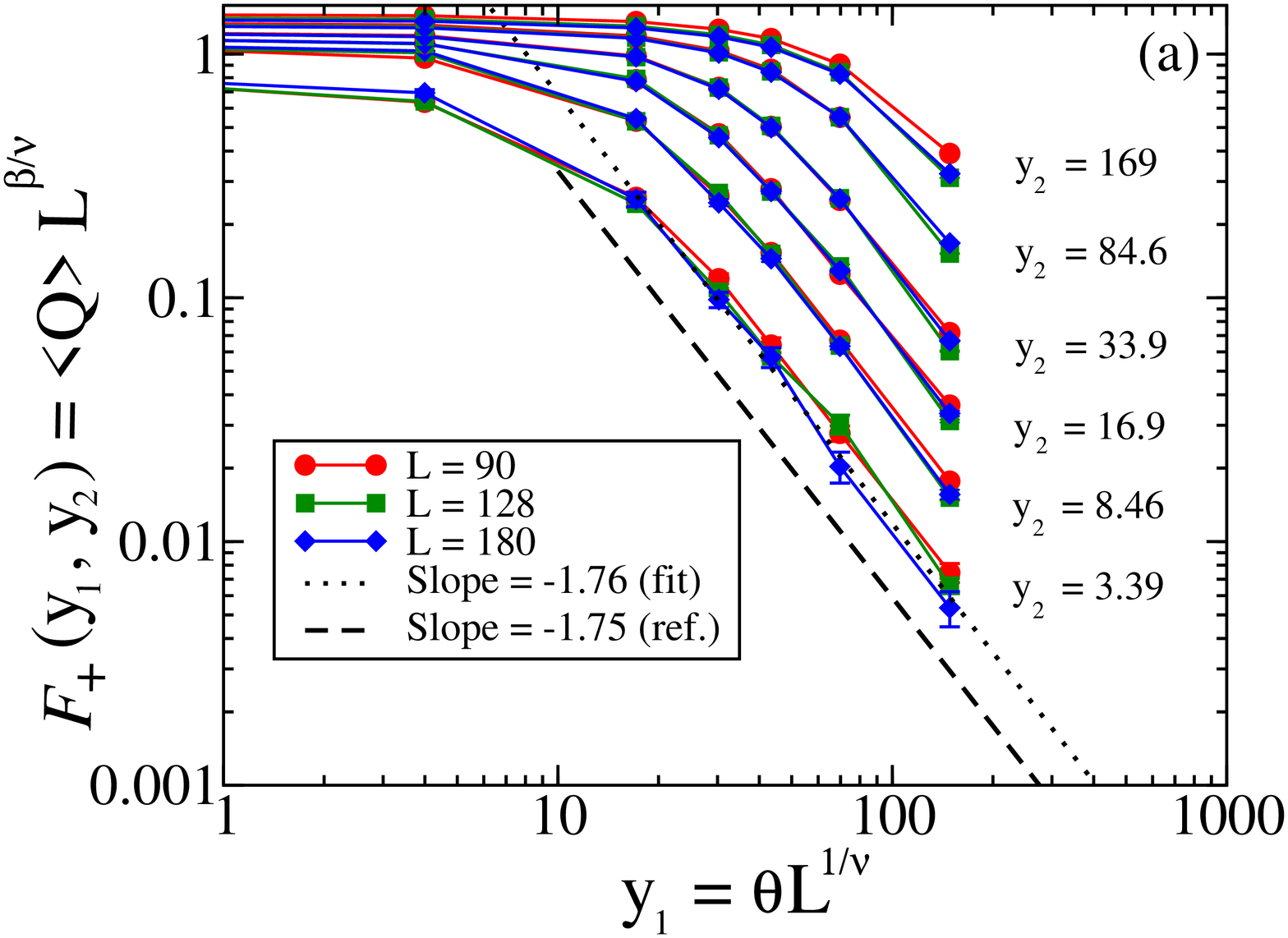}  
\includegraphics[width=3.5in,angle=270]{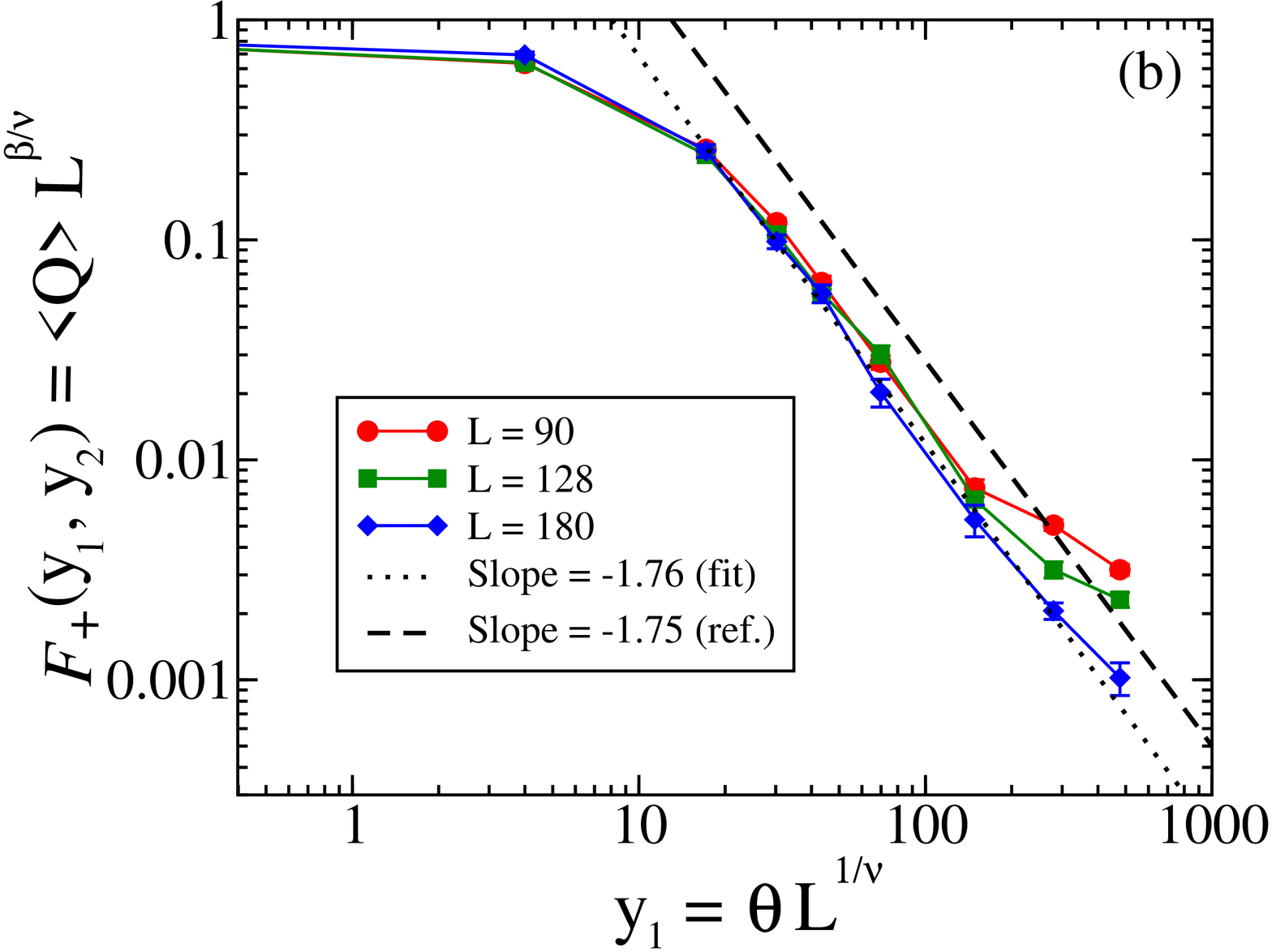} 
\end{center}
\caption
{ 
(Color online.) Log-log plots of the scaling function $\mathcal{F}_+(y_1,y_2)$ vs $y_1$ for lattice
sizes $L=90, 128,$ and $180$. The values of $y_1$ plotted are 4.00, 17.2, 30.3, 43.4, 69.7, 147, and
(in \textbf{(b)}) 280 and 477. \textbf{(a)} The data are shown for the values of $y_2$ listed on the plot.
The best-fit line for the last five points of the $L = 180$ data at $y_2 = 3.39$, with slope
$-1.76 \pm 0.07$, is included along with a reference line with the slope $-\gamma_{\mathrm{e}} = -1.75$. \textbf{(b)}
The data for $y_2 = 3.39$ with two additional $y_1$ values illustrates the boundary of the regime of power-law scaling.
}
\label{f+.vs.y1}
\end{figure}

In Fig.~\ref{f+.vs.y1}(a), using the equilibrium values $\beta_{\mathrm{e}}$
and $\nu_{\mathrm{e}}$ in calculating 
$\mathcal{F}_+(y_1,y_2) \equiv L^{\beta / \nu} \langle Q \rangle_L$,  we 
present a plot of the scaling function $\mathcal{F}_+$ vs $y_1$ for 
different values of $y_2$, for lattice sizes $L = 90$, 128, and 180. 
Here and for the remainder of the paper, exponents with the subscripts `d' and
`e' refer to the behavior of the nonequilibrium system (with a dynamic phase
transition) and the equilibrium system, respectively. The scaling function exhibits
a power-law dependence in the regime $y_1 \gg y_2$, which is consistent with Eq. (\ref{eq:Fscal}). At progressively
larger values of the constant $y_2$, the power-law scaling can be seen to begin at increasing values of
$y_1$, as would be expected. A best-fit line to the final five points of the $L = 180$ data at $y_2 = 3.39$
yields an estimate of the scaling exponent  
$-\gamma_{\mathrm{d}} = -1.76 \pm 0.07$ in  Eq. (\ref{eq:Fscal}). 
This is consistent with the previous results for $H_c = 0$ cited above
\cite{kn:korniss01}, and it supports the hypothesis
that $\gamma_{\mathrm{d}} = \gamma_{\mathrm{e}} = 7/4 = 1.75$.  In Fig.~\ref{f+.vs.y1}(b), 
we present just the data for $y_2 = 3.39$, including
additional data points at $y_1 = 280$ and $477$. The data deviate from the
power-law behavior for $L=90$ at
$y_1 > 149$, for $L=128$ at $y_1 > 280$, and for $L=180$ at $y_1 > 477$. This locates the boundary
of the scaling regime (for $y_2 = 3.39$) at $ \theta = y_1 / L^{1/\nu} \approx  2.65$ .

\begin{figure}[ht]
\begin{center}
\includegraphics[width=4.5in,angle=270]{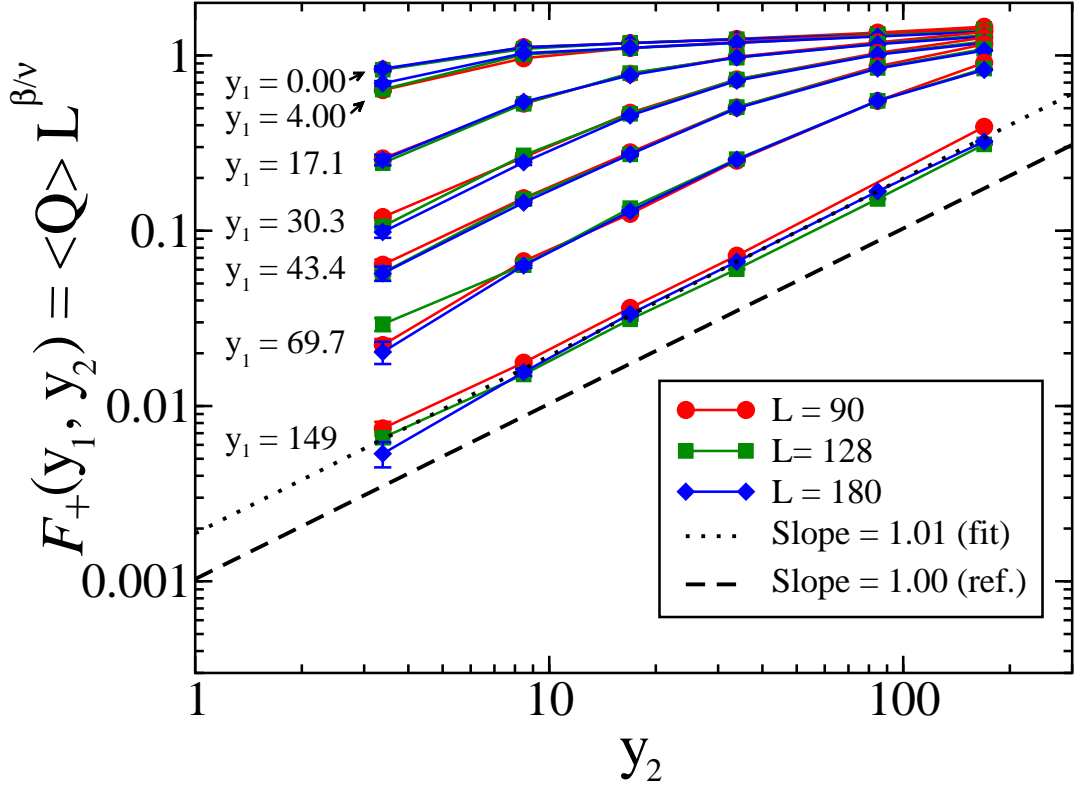} 
\end{center}
\caption
{ 
(Color online.) Log-log plot of the scaling function $\mathcal{F}_+(y_1,y_2)$ vs $y_2 = (H_b/J) L^{\beta \delta / \nu}$ for lattice sizes $L = 90,$
128, and 180, for the constant values of $y_1$ labeled in the plot. The values
of $y_2$ used are $y_2 = 3.39, 8.46, 16.9, 33.9, 84.6,$ and $169$. The dotted line represents the best fit
to the first five points of the $L = 180$ data at $y_1 = 149$, and has a slope 
of $1.01 \pm 0.01$. The dashed line shows the slope value of 1, expected from Eq.
(\ref{eq:Fscal}).
}
\label{f+.vs.y2.smally2} 
\end{figure}

In Fig.~\ref{f+.vs.y2.smally2}, again using $\beta_{\mathrm{e}}$
and $\nu_{\mathrm{e}}$ in calculating $\mathcal{F}_+$, we plot the scaling 
function $\mathcal{F}_+$ vs $y_2$ at different values of $y_1$, in order
to examine the scaling behavior for $y_1 \gg y_2$. For the constant values $y_1 = 43.4, 69.7,$
and 149, power-law scaling can be observed in the regime $y_1 \gg y_2$. 
A best-fit line to the $y_1 = 149$ data
for the five points from $y_2 = 3.39$ to $84.6$ yields a scaling exponent of $1.01 \pm 0.01$, which
is consistent with the value of $1$ expected from Eq. (\ref{eq:Fscal}).

\begin{figure}[ht]
\begin{center}
\includegraphics[width=4.5in,angle=270]{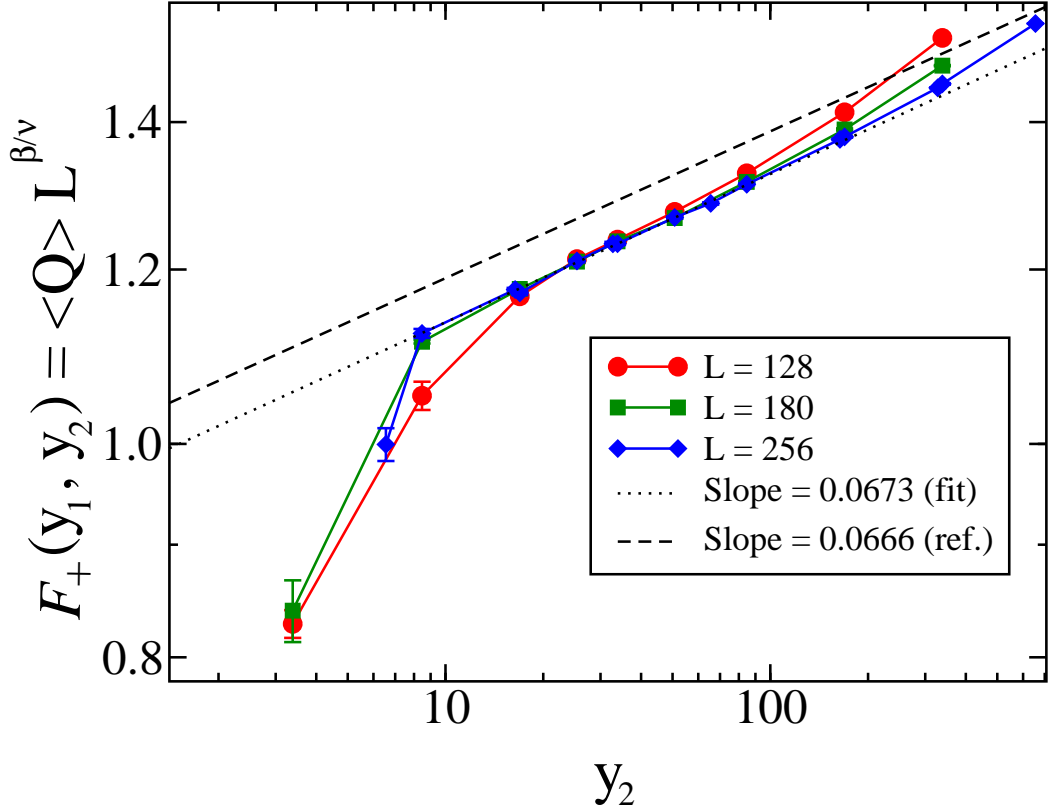} 
\end{center}
\caption
{ 
(Color online.) Log-log plot of the scaling function  $\mathcal{F}_+(y_1,y_2)$ vs $y_2 = (H_b/J)  L^{\beta \delta / \nu}$ for lattice 
sizes $L = 128,$ 180, and 256, at the critical period $P_c$, where $y_1 = 0$. 
In the $L = 256$ data, near the values $y_1 = 16, 32, 165,$ and 335, two
closely spaced data points are actually plotted. The best-fit line to the $L=256$ data 
in the range $8.46 < y_2 < 84.6$, shown as a dotted line in the plot, 
corresponds to a scaling exponent $1/\delta_{\mathrm{d}} = 0.0673 \pm 0.0008$. A reference line corresponding to scaling exponent
$1 / \delta_{\mathrm{e}} = 1/ 15 = 0.0666$ is also shown. These results are in complete agreement with those shown in Fig. \protect\ref{delta.scaling}.
}
\label{f+.vs.y2.largey2} 
\end{figure}

In order to investigate the scaling of $\mathcal{F}_+$ in the asymptotic 
limit $y_1 \ll y_2$, we plot in Fig.~\ref{f+.vs.y2.largey2} the 
scaling function 
$\mathcal{F}_+(y_1,y_2)$ vs $y_2$ at the critical period $P=P_c$ 
(i.e., $y_1 = 0$), at lattice sizes $L = 128, 180,$ and 256. 
In the range $20 < y_2 < 50$, 
power-law scaling is observed for all three lattice sizes. For $y_2 > 50$, the data deviate from
power-law scaling, with the smallest lattice size deviating first, as expected
in a finite-size scaling plot.  A fit of the $L=256$ data 
in the range from $y_2 = 8.46$ to $84.6$
produces a scaling exponent $0.0673 \pm 0.0008$. 
Since the constant factors $L^{\beta \delta / \nu}$
and $L^{\beta / \nu }$ do not affect the fit of the scaling exponent, this
is the same exponent found in the fit of $\langle Q \rangle$ vs $H_b$ in 
Fig.~\ref{delta.scaling}.
The reciprocal of this scaling exponent is thus 
$\delta_{\mathrm{d}} = 14.85 \pm 0.18$, which is consistent with the exponent of the critical isotherm, $\delta_{\mathrm{e}} = 15$,
in the equilibrium Ising model.

The comparison of the second scaling function, $\mathcal{G}_+(y_1,y_2$), 
to numerical data is more clearly presented after the relationship of 
the susceptibility $\hat{\chi}_L$ and the scaled variance $X_L^Q$ has been examined. 
Therefore, we present in the next section numerical
results on the extent of applicability of an FDR between $\hat{\chi}_L$ 
and $X_L^Q$, before turning in Sec.~\ref{s:second.compare} 
to the second scaling function.

\section{Applicability of a fluctuation-dissipation relation}
\label{s:fluct.relation}

FDRs, such as the Einstein relation, Green-Kubo relations, etc., hold a
central place in equilibrium statistical mechanics. This is essentially a
consequence of detailed balance and
the role of the partition function as a moment-generating
function, and thus such relations cannot be readily extended to
nonequilibrium steady states. However, it has recently been shown that
certain FDRs can be extended to far-from equilibrium steady states by use of
an {\it effective temperature\/} \cite{HAYA04,HAYA05}. 
Here we will therefore consider whether the nonequilibrium susceptibility and
the scaled variance of the dynamic order parameter can be related as 
\begin{equation}
\hat{\chi}_L \equiv \frac{\partial \langle Q \rangle_L}{\partial H_b} = 
\frac{ L^2 \left( \langle Q^2 \rangle_L - \langle Q \rangle_L^2 \right) } 
{T_{\mathrm{eff}}} \equiv \frac{X_L^Q}{T_{\mathrm{eff}}},
\label{eq:FDR}
\end{equation}
with an effective temperature $T_{\mathrm{eff}}$, 
in a way analogous to the equilibrium FDR, 
\begin{equation}
\chi_L \equiv \frac{\partial \langle m \rangle_L}{\partial H} 
= \frac{ L^2 \left( \langle m^2 \rangle_L - \langle m \rangle_L^2 \right)}{T}
\;,
\label{eq:FDR0}
\end{equation}
in which $T$ is the temperature. 
As mentioned in Secs.~\ref{s:intro} and~\ref{s:finite.size}, 
this conjecture motivated the use in previous work of the
scaled variance $X_L^Q$
as a proxy for $\hat{\chi}_L$ in investigating the scaling 
behavior of the nonequilibrium system near its critical period. 

\begin{figure}[ht]
\begin{center}
\includegraphics[width=4.5in, angle=270]{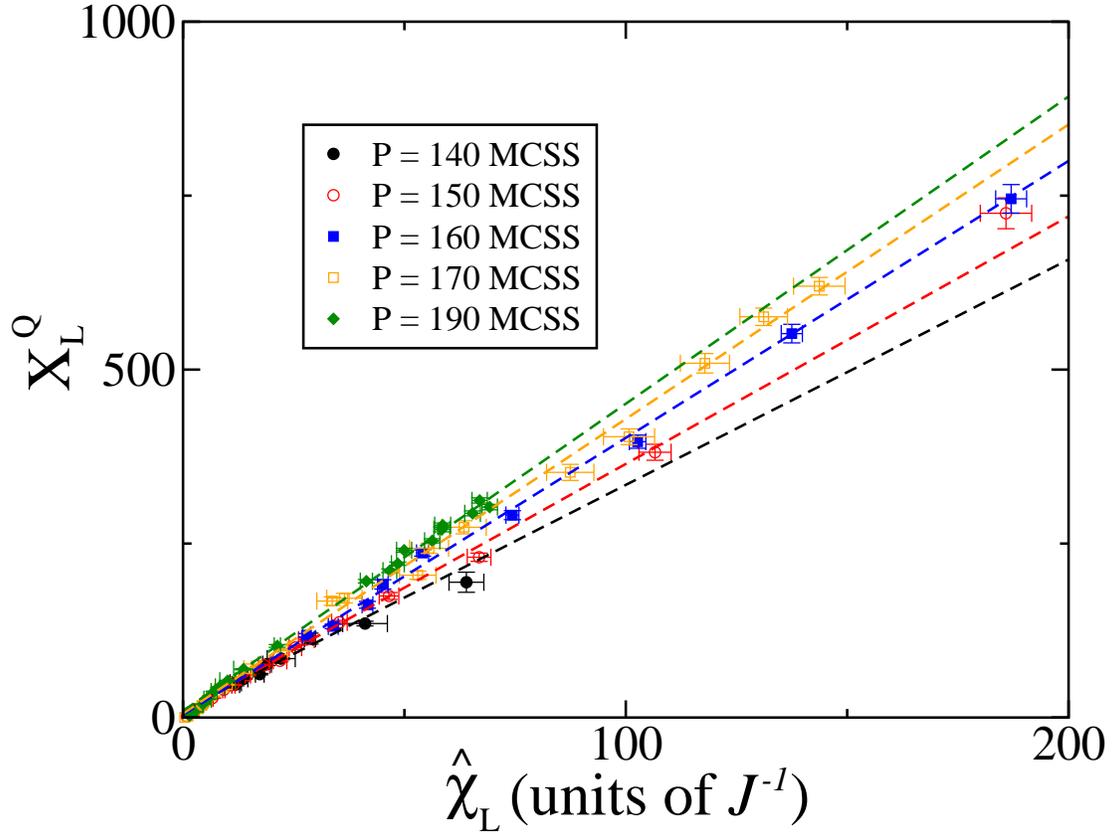}
\end{center}
\caption{ 
(Color online.) The scaled fluctuations $X_L^Q$ of the dynamic order paramater plotted vs 
its susceptibility $\hat{\chi}_L$ to the bias field $H_b$, calculated at $L=180$, for
periods $P = 140, 150, 160, 170$ and 190 MCSS. The quantity $\hat{\chi}_L$ was
calculated using the numerical derivative in Eq. (\ref{eq:num.deriv}). The best-fit lines
shown, whose slopes increase monotically with the period $P$ of the data to which they were
fit, were calculated as 
$(3.239J)\hat{\chi}_L + 10.42$, $(3.557J)\hat{\chi}_L + 8.735$,  
$(3.980J)\hat{\chi}_L + 3.889$,  $(4.232J)\hat{\chi}_L + 5.868$ and  
$(4.497J)\hat{\chi}_L + 5.371$, respectively.
}
\label{fluct.susc} 
\end{figure}

\begin{figure}[ht]
\begin{center}
\includegraphics[width=4.5in, angle=270]{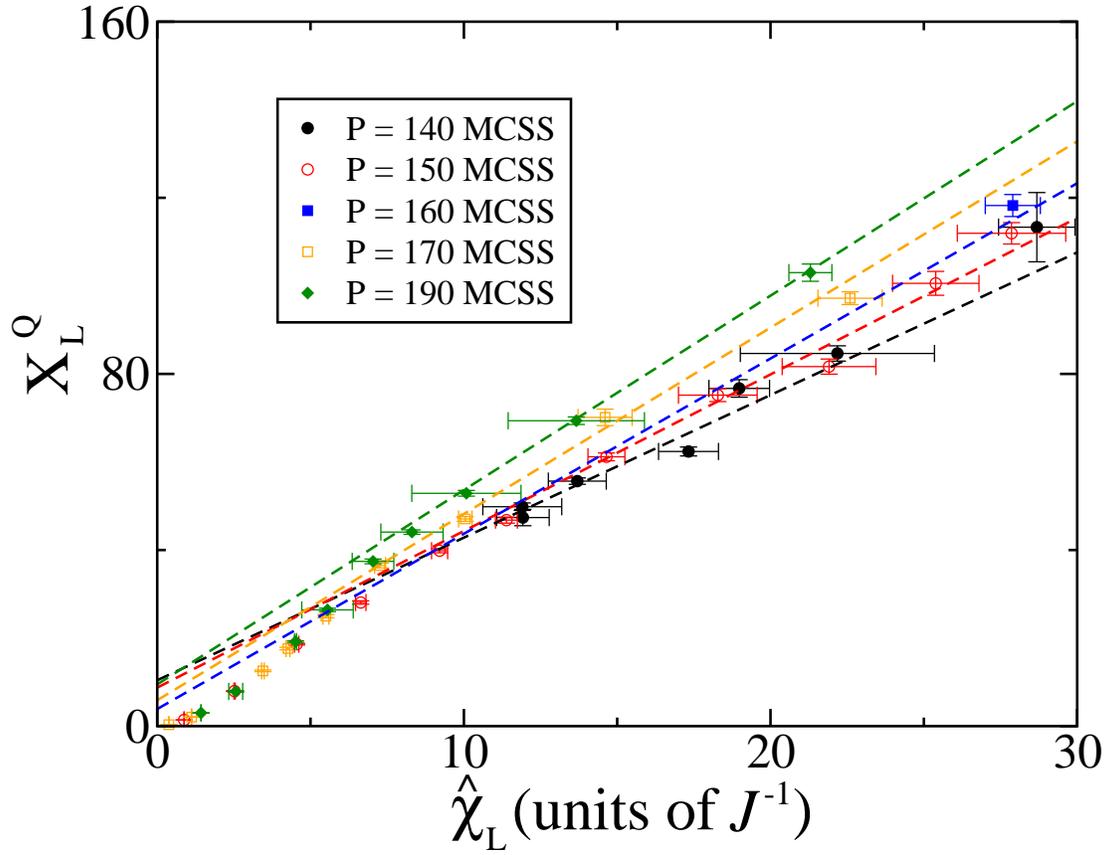}
\end{center}
\caption{ 
(Color online.) Closeup of Fig.~\protect\ref{fluct.susc}, showing the relationship of $X_L^Q$ and
$\hat{\chi}_L$ at low values of $\hat{\chi}_L$, which correspond to large values of the
bias field $H_b$. For $P = 150$, 170 and 190 MCSS, data have been taken (and are shown) down to 
very low values of $\hat{\chi}_L$, where the breakdown of the linear relationship 
between  $X_L^Q$ and $\hat{\chi}_L$ can be clearly seen. The dashed lines are 
the same best-fit lines shown in Fig. \ref{fluct.susc}.
}
\label{fluct.susc.lowchi} 
\end{figure}

To test the extent to which Eq.~(\ref{eq:FDR}) holds, we
computed values of $\hat{\chi}_L$ and $X_L^Q$ for a 
range of periods from $P=140$ to 250~MCSS and a range of bias fields from  
$H_b = 0$ to an upper limit between $0.005J$ and $0.2J$. 
(The bias field necessary to `saturate' the nonequilibrium system,
i.e., to produce values of $\hat{\chi}_L$ and $X_L^Q$ near zero, increases as 
the period is increased.) The computations were perfomed at $L = 180$.
The quantity $\hat{\chi}_L$ was computed directly
as a numerical derivative:
\begin{equation}
\hat{\chi}_L (P,H_b) \approx \left(\langle Q \rangle(P,H_b+\Delta H_b) - \langle
Q \rangle (P,H_b-\Delta H_b)\right) / 2\Delta H_b .
\label{eq:num.deriv}
\end{equation}
The choice of $\Delta H_b = 0.1 H_b$ was found to produce
sufficiently accurate values of the numerical derivative
across the range of bias fields studied.
The results for periods $P = 140$~through 190 MCSS are shown in
Fig.~\ref{fluct.susc}.  A linear relationship is seen to
exist between $\hat{\chi}_L$ and $X_L^Q$, for each value of $P$, over a wide
range of $\hat{\chi}_L$ values. 
At each period, the dependence becomes nonlinear below a certain value of 
 $\hat{\chi}_L$, as illustrated for periods $P=150$, 170, and 190 MCSS 
in Fig.~\ref{fluct.susc.lowchi}. Since low values of the susceptibility $\hat{\chi}_L$
correspond to large values of the bias field $H_b$, we interpret this breakdown
of linearity as an indication that the FDR in Eq. (\ref{eq:FDR}) holds only in a
scaling regime around the critical point, i.e., for a limited range of $H_b$ around $H_b=0$.

\begin{figure}[ht]
\begin{center}
\includegraphics[width=4.5in, angle=270]{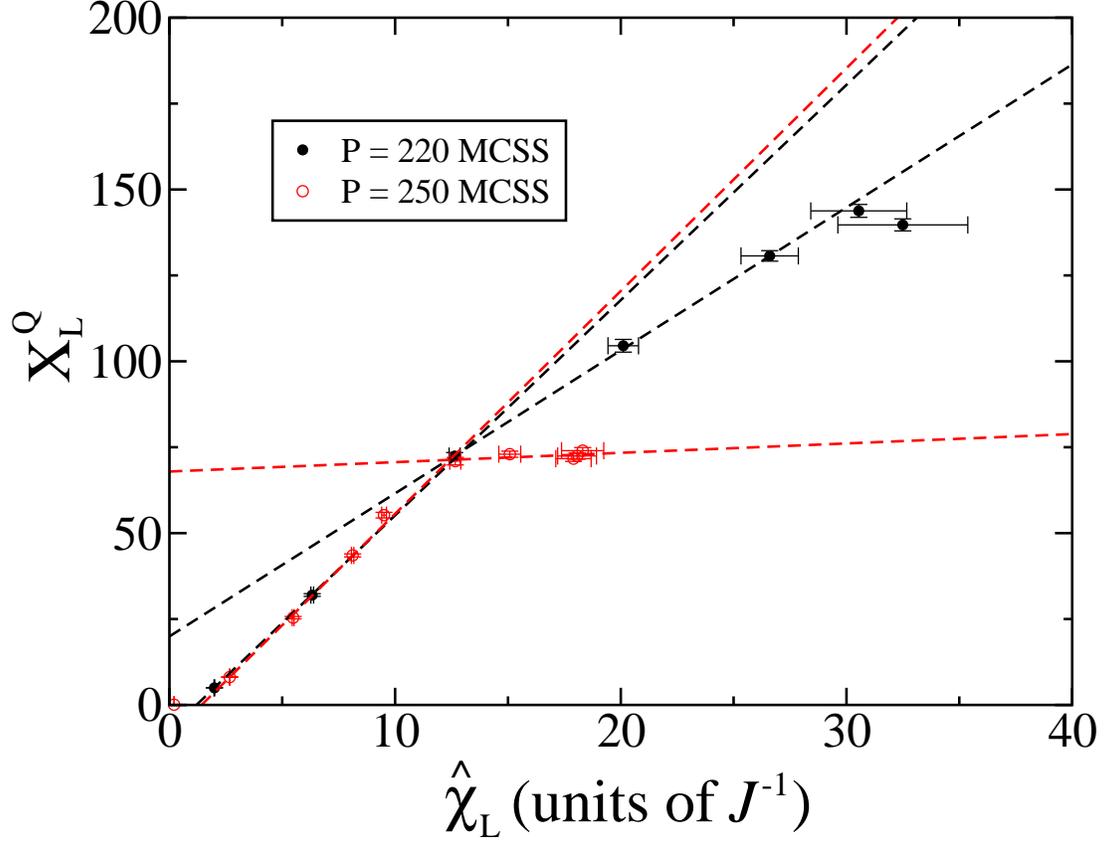}
\end{center}
\caption{
(Color online.) 
The scaled fluctuations $X_L^Q$ of the dynamic order paramater plotted vs
its susceptibility $\hat{\chi}_L$ to the bias field $H_b$, calculated for lattice size
 $L=180$, at 
periods $P = 220$ and 250 MCSS. At each period, the data were fit (purely phenomenologically) 
to two linear relationships. For $P = 220$~MCSS, the fits were calculated as
$(6.265J)\hat{\chi}_L - 7.497$ at low $\hat{\chi}_L$, and
$(4.161J)\hat{\chi}_L + 19.94$ at high $\hat{\chi}_L$.  For $P = 250$~MCSS,
the fits were $(6.485J)\hat{\chi}_L - 9.240$ at low $\hat{\chi}_L$, and   
$(0.2726J)\hat{\chi}_L + 67.92$ at high $\hat{\chi}_L$. 
}
\label{fluct.susc2} 
\end{figure}

The relationship between $X_L^Q$ and $\hat{\chi}_L$ at the higher periods,
$P = 220$ and $P = 250$~MCSS, is more complicated, as shown in Fig.~\ref{fluct.susc2}. 
At $P = 220$~MCSS, following the nonlinear regime at low
$\hat{\chi}_L$, there is a linear relationship with slope 
$T_{\mathrm{eff}} = (6.27 \pm 0.11) J$ up to $\hat{\chi}_L \approx 13~J^{-1}$, 
followed 
by a second distinct linear dependence with slope $T_{\mathrm{eff}} 
\approx (4.16 \pm 0.29) J$ above
$\hat{\chi}_L = 13~J^{-1}$. 
At $P = 250$~MCSS, the initial nonlinear dependence is
again present. Then the first linear regime has 
$T_{\mathrm{eff}} = (6.49 \pm 0.07) J$ up to $\hat{\chi}_L \approx 13~J^{-1}$, 
and is followed by a regime which can be characterized as either linear 
with very gentle slope $(0.27 \pm 0.22) J$,
or as an effective `saturation' of $X_L^Q$ past $\hat{\chi}_L = 13~J^{-1}$. 

\begin{figure}[ht]
\begin{center}
\includegraphics[width=4.5in,angle=270]{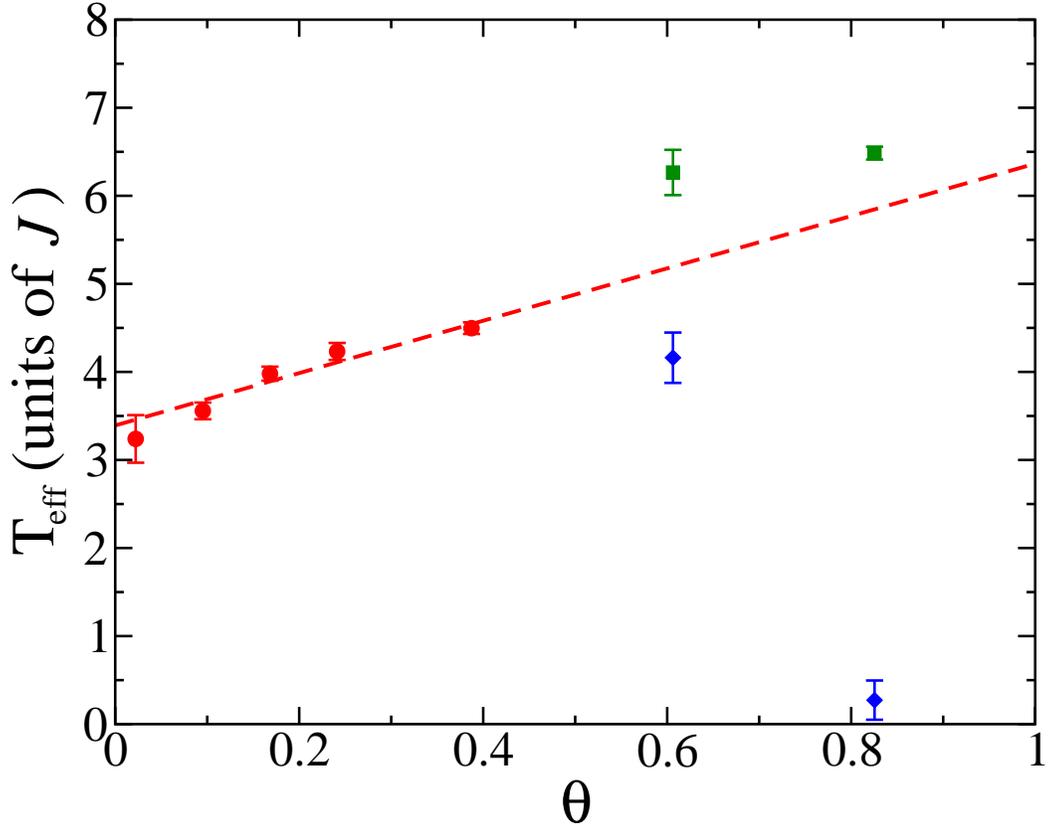}
\end{center}
\caption{
(Color online.) 
The effective temperature $T_\mathrm{eff}$, obtained as the slopes of the 
linear fits to the data in Figs.~\protect\ref{fluct.susc} and 
\protect\ref{fluct.susc2}, plotted vs 
$\theta = \left( P - P_c \right) / P_c$. For the values $\theta = 0.606$ and 
0.825 ($P = 220$ and 250 MCSS), the slopes of both linear regimes fit in
Fig.~\protect\ref{fluct.susc2} are plotted as $T_\mathrm{eff}$ values, using
filled squares and diamonds rather than filled circles. 
The straight line is a weighted least-squares fit to
the data below $\theta \approx 0.4$ ($P<190$~MCSS), and has a slope of $2.97J$.
}
\label{teff.theta} 
\end{figure}

In Fig.~\ref{teff.theta} we plot the best-fit slopes from Figs.~\ref{fluct.susc}
and \ref{fluct.susc2}, which according to
Eq. (\ref{eq:FDR}) represent estimates of $T_{\mathrm{eff}}$, vs the
scaling parameter $\theta = \left( P-P_c \right) / P_c $. 
We have included in the plot the slopes of both linear regimes for the values 
$\theta = 0.606$ and 0.825 ($P = 220$ and 250 MCSS). For $\theta$ below 0.4
($P \approx 190$~MCSS), $T_{\mathrm{eff}}$ increases with
$\theta$ in a way not inconsistent with a linear relationship (with slope $2.97J$). It may be 
interesting to note that an extrapolation of the linear relationship to 
$\theta = 0$ ($P = P_c$) yields the value 
$T_{\mathrm{eff}} = 3.39J$, which is significantly higher than 
the critical temperature, $T_c = 2.2619J$, of the 
equilibrium Ising system. However, one should not put too much emphasis on the numerical values
of $T_{\mathrm{eff}}$ \cite{eff.temp.endnote}, as they could easily be changed. For instance, if $H_b$ is only
{\it proportional to} the full conjugate field $H_c$, with a proportionality constant
different from unity, this would trivially change $T_{\mathrm{eff}}$ in Eq.~(\ref{eq:FDR}).
The important result, which we have demonstrated to hold in the critical region, is the linear relationship between $X_L^Q$ and $\hat{\chi}_L$.  

We can thus characterize
the extent of applicability of an FDR to the DPT above the critical period as
follows.  For $\theta < 0.4$, an FDR holds outside of a small nonlinear regime
at low $\hat{\chi}_L$ (high $H_b$), with an effective temperature 
$T_{\mathrm{eff}}$ which increases approximately linearly with $\theta$. For $\theta$ above 0.4, 
two linear relationships appear to exist between $X_L^Q$ and $\hat{\chi}_L$ 
in separate regimes, making it impossible to define a unique $T_{\mathrm{eff}}$
at a given value of $\theta$.  An understanding of the nonlinear regime, which
is present at low 
$\hat{\chi}_L$ for all periods examined, as well as of the complication 
of the FDR above $\theta = 0.4$, would be highly desirable.
We hope that these numerical results can stimulate the
development of, as well as test the accuracy of, a theoretical description of the
non-equilibrium steady states produced in the presence of non-zero $H_b$ 
for this DPT.

\section{Comparison of second scaling function to computational results}
\label{s:second.compare}

\begin{figure}[ht]
\begin{center}
\includegraphics[width=4.5in,angle=270]{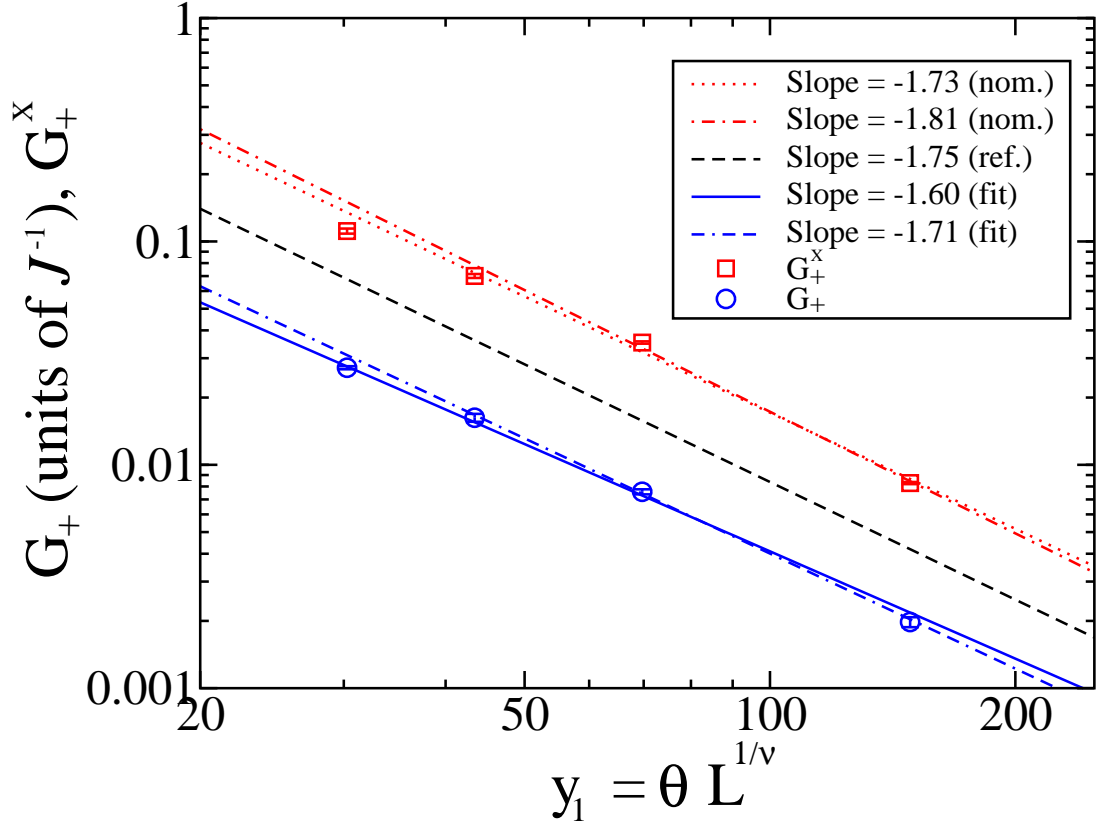} 
\end{center}
\caption
{ (Color online.)  
Log-log plots of $\mathcal{G}_+(y_1,y_2)$ and $\mathcal{G}_+^X(y_1,y_2)$ vs $y_1$, over
the range $y_1 = 30.3$ through 149, for $y_2 = 8.46$ at $L=180$. 
The relatively small error bars on each data point 
can be seen inside the larger symbols.  
The solid and dash-dash-dotted lines 
are fits to all four and the last three $\mathcal{G}_+$ data points, respectively, and
correspond to scaling exponents $-1.60 \pm 0.03$ and $-1.71 \pm 0.05$. The dotted and
dash-dotted lines are the result of attempts to fit all four and the last three  
$\mathcal{G}_+^X$ data points, respectively. 
They correspond to nominal scaling exponents $-1.73 \pm 0.01$ and
$-1.81 \pm 0.02$. The dashed line is a reference line corresponding to scaling exponent $-1.75$.
}
\label{gvsy1.h2}
\end{figure}

\begin{figure}[ht]
\begin{center}
\includegraphics[width=4.5in,angle=270]{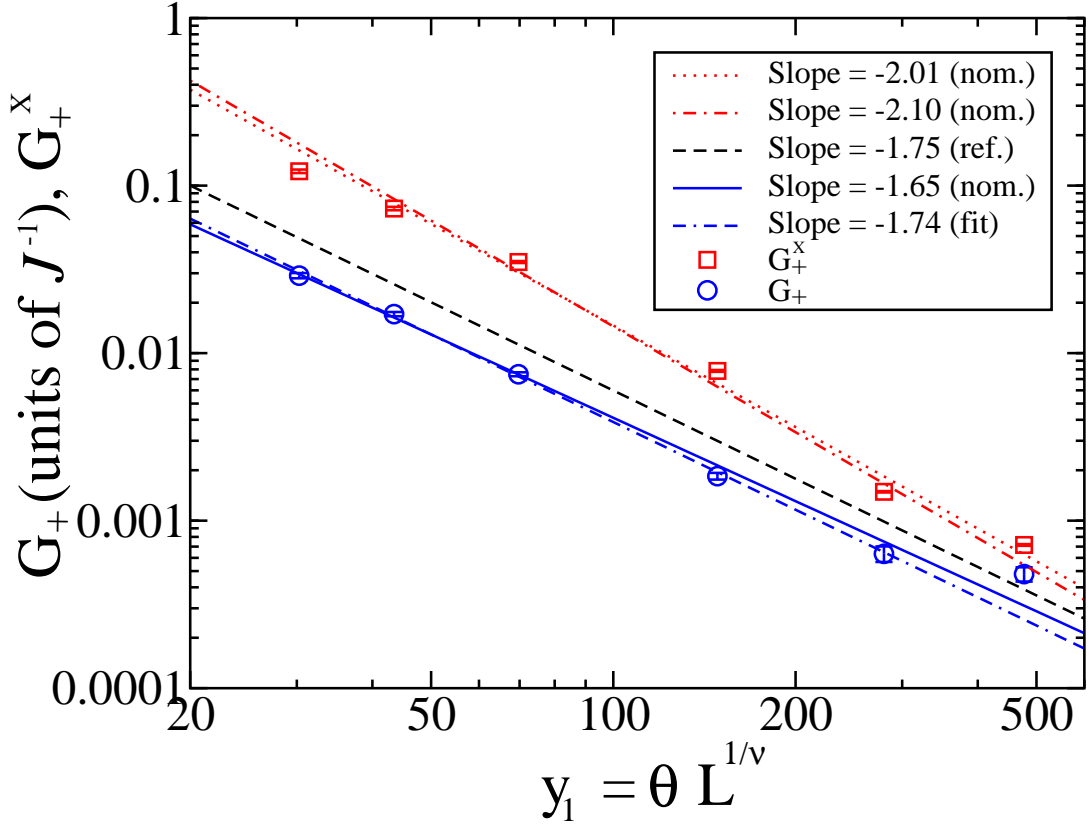} 
\end{center}
\caption
{ 
(Color online.) 
Log-log plots of $\mathcal{G}_+(y_1,y_2)$ and $\mathcal{G}_+^X(y_1,y_2)$ vs $y_1$, over
the range $y_1 = 30.3$ through 477, for $y_2 = 0$ at lattice size $L=180$. 
The relatively small error bars on each data point 
can be seen inside the larger symbols.  
The solid and dash-dash-dotted lines show the result of attempts to fit
all six and the first five $\mathcal{G}_+$ data points, and
correspond to a nominal scaling exponent $-1.65 \pm 0.03$, and a statistically
significant scaling exponent $-1.74 \pm 0.03$, respectively. The dotted and
dash-dotted lines are the result of attempts to fit all six and the first 
five $\mathcal{G}_+^X$ data points, and correspond to nominal scaling exponents
 $-2.01 \pm 0.01$ and $-2.10 \pm 0.01$, respectively. The dashed line is a 
reference line corresponding to scaling exponent $-1.75$.
}
\label{gvsy1.h0}
\end{figure}

\begin{figure}[ht]
\begin{center}
\includegraphics[width=4.5in,angle=270]{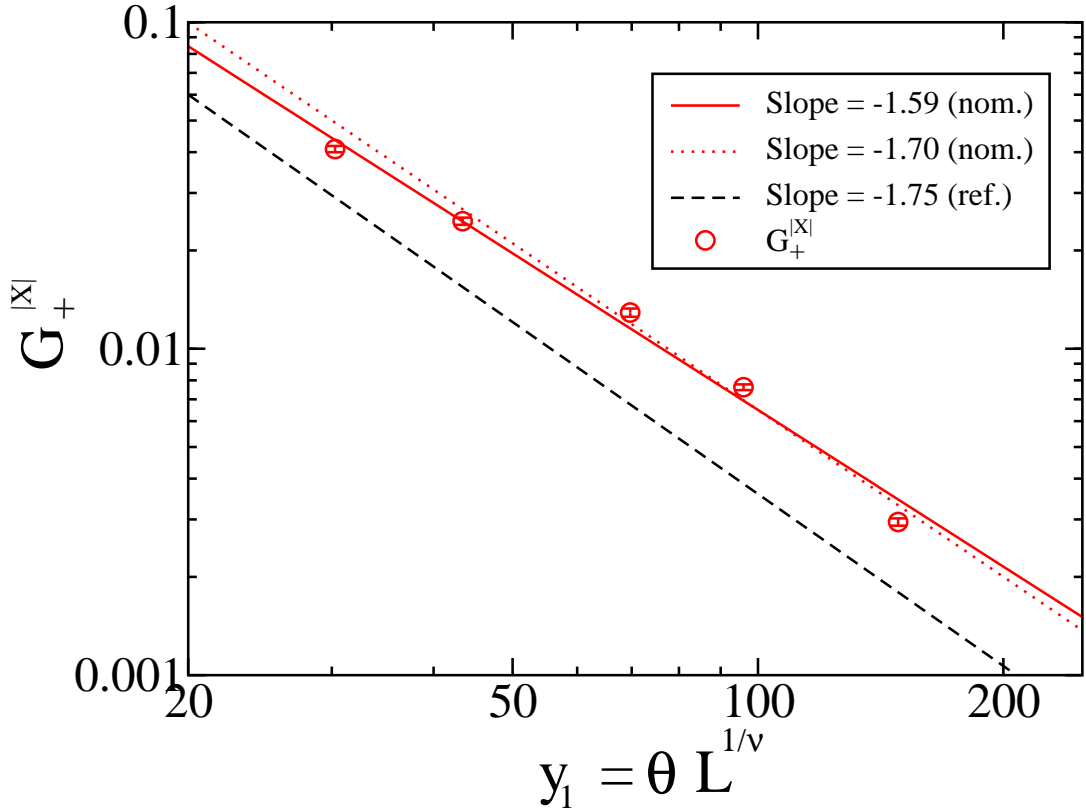} 
\end{center}
\caption
{ 
(Color online.)  
Log-log plot of $\mathcal{G}_+^{|X|}(y_1,y_2)$ vs $y_1$, over
the range $y_1 = 30.3$ through 149, for $y_2 = 0$ at lattice size $L=180$.
The relatively small error bars on each data point 
can be seen inside the larger symbols.  
The solid and dotted lines are the results of attempts to fit all five and the 
last four data points with a power-law relationship, and correspond to 
nominal scaling exponents of $-1.59 \pm 0.02$ and $-1.70 \pm 0.03$, 
respectively. The dashed line is a reference line corresponding to a scaling
exponent of $-1.75$.
}
\label{gvsy1.h0.xabsq}
\end{figure}

We now test the asymptotic scaling forms for $\mathcal{G}_+$ in Eq. 
(\ref{eq:Gscal}). First, we note that in performing least-squares fits, 
one normally requires the
goodness-of-fit parameter $q$, i.e., the probability that (assuming the fit
relationship were true) random error alone could produce the observed data, to
be greater than $10^{-3}$ to consider the fit reasonable. Within this
section, however, and in the captions to Figs.~\ref{gvsy1.h2} through 
\ref{gvsy1.h0.xabsq}, it will be useful for descriptive purposes to refer to
scaling exponents resulting from attempts at least-squares fits with $q$ 
values below this acceptable range. We will refer to the results of such
unsuccessful fitting attempts as `nominal' scaling exponents, and for clarity
will report the value of the parameter $q$ for each scaling exponent presented
in this section.

In Fig.~\ref{gvsy1.h2}, we
plot the scaling function $\mathcal{G}_+$ vs $y_1$ for 
$y_2 = 8.46$ at $L=180$, using $\gamma_{\mathrm{e}}$ and $\nu_{\mathrm{e}}$ to
calculate $\mathcal{G}_+$, and evaluating $\hat{\chi}_L$ numerically
according to Eq. (\ref{eq:num.deriv}). In addition, we plot in the same figure
the scaling function $\mathcal{G}^X_+(y_1,y_2) \equiv X_L^Q L^{-\gamma/\nu}$ vs
$y_1$, for the same values of $y_2$ and $L$.
A fit to all four $\mathcal{G}_+$ data points yields a scaling exponent
$-1.60 \pm 0.03$ ($q = 0.02$), while a fit to the 
last three $\mathcal{G}_+$ data points
yields a scaling exponent $-1.71 \pm 0.05$ ($q = 0.25$). 
We will provide evidence in the next paragraph that only the last 
three $\mathcal{G}_+$ data 
points, and not the first, satisfy the asymptotic condition $y_1 \gg y_2$. 
Thus, these data are consistent with power-law scaling of $\hat{\chi}_L$ with exponent
$-\gamma_{\mathrm{d}} = -\gamma_{\mathrm{e}} = -7/4 = -1.75$. Attempts to fit the
$\mathcal{G}^X_+$ data to all four and the last three data points yield 
nominal scaling exponents $-1.73 \pm 0.01$ ($q < 10^{-15}$) and 
$-1.81 \pm 0.02$ ($q = 2.3 \times 10^{-14}$), respectively. Thus, while
in Fig.~\ref{gvsy1.h2} it appears that the power-law relationships with these
nominal scaling exponents give respectable visual fits to the $\mathcal{G}^X_+$
data, there are variations in the data which, while small, are larger than the 
statistical error bars, and which prevent
a statistically significant fit. We will describe the causes of these 
variations later in this section.

We present in Fig.~\ref{gvsy1.h0} a plot of
$\mathcal{G}_+$ and $\mathcal{G}^X_+$ vs $y_1$ at $y_2 = 0$,
again for $L=180$, for a larger range from $y_1 = 30.3$ to 477.
With $y_2 = 0$, we expect that $y_1 = 30.3$ (and indeed, essentially 
any nonzero value
of $y_1$) should satisfy the asymptotic scaling condition $y_1 \gg y_2$.
An attempt to fit to all six $\mathcal{G}_+$ data points in Fig.~\ref{gvsy1.h0} gives a nominal scaling exponent $-1.65 \pm 0.03$ 
($q = 3.8 \times 10^{-7}$), while a fit to the first five points
($y_1 = 30.3$ through 280) gives a scaling exponent $-1.74 \pm 0.03$  
($q=0.07$). Excluding the first data point at $y_1 = 30.3$ has little effect
on either fit. This supports the assumption that with $y_2 = 0$, 
the asymptotic scaling condition
$y_1 \gg y_2$ holds for $y_1 = 30.3$, while for $y_2 = 8.46$, as used in 
Fig.~\ref{gvsy1.h2}, the asymptotic scaling condition does not hold for $y_1 = 30.3$.
Attempts to fit all of, and the first five of, the $\mathcal{G}^X_+$ data 
points to power-law scaling again yield only nominal scaling exponents 
$-2.01 \pm 0.01$ 
($q < 10^{-15}$) and $-2.10 \pm 0.01$ ($q < 10^{-15}$), respectively. 

We considered that the low statistical significance of the fits to the
$\mathcal{G}^X_+$ data could be caused by underestimation of the 
error bars on $X_L^Q$. These error bars were calculated by
(i) finding the correlation time in the numerical data series $Q_i$ from
the simulation, and sampling data at intervals of twice the correlation time;
(ii) dividing this sampled data into $k>16$ groups and calculating the value
of $X_L^Q$ within each group; (iii) finding the mean and standard error of
this collection of $X_L^Q$ values. As a check on self-consistency, we
performed several independent calculations of $X_L^Q$ by this method, and
found that the standard error of these values (corrected for small sample
size) was comparable to the standard error found within each calculation.
Thus, we have strong evidence that the error bars for $X_L^Q$
(and $\mathcal{G}^X_+$) are accurate.

These scaling results can be understood in light of the 
observations in Sec.~\ref{s:fluct.relation} on the relationship between
$X_L^Q$ and $\hat{\chi}_L$. We can reasonably assume that the breakdown in 
scaling of $\mathcal{G}_+$ past $y_1 = 280$ ($P = 350$~MCSS, $\theta = 1.56$)
in Fig.~\ref{gvsy1.h0} occurs because this is the boundary of the critical region. The 
small variations of the $\mathcal{G}^X_+$ data for $P < 350$~MCSS 
in Fig.~\ref{gvsy1.h0} from a scaling relationship with exponent 
$-\delta_{\mathrm{d}} = -1.75$ then have three main causes.
The first cause is the multiplication of the
accurately scaling function $\mathcal{G}_+$ by the $\theta$- and $y_1$-dependent
value $T_{\mathrm{eff}}$, according to Eq. (\ref{eq:FDR}).  However, such a
variation would also occur in an analogous plot for the scaling 
of  $X_L^M \equiv L^2 \left(\langle m^2 \rangle -
\langle m \rangle ^2 \right)$ vs
$y_{1,\mathrm{e}} = \epsilon L^{1/\nu} = \left(\left(T-T_c\right)/T_c\right)L^{1/\nu}$ 
in the equilibrium Ising model, since the susceptibility 
$\chi_L^M \equiv \partial \langle m \rangle / \partial H$ scales with exponent
$-\gamma_{\mathrm{e}}$, and $X_L^M$ is related to $\chi_L^M$ by the 
$\epsilon$-dependent temperature $T$ according to Eq. (\ref{eq:FDR0}).
This effect is small enough to be neglected in equilibrium critical scaling, 
and, since the change in $T_{\mathrm{eff}}$ from $\theta = 0.02$ to 
$\theta = 0.4$ is comparable to the change in $T$ from $\epsilon = 0.02$ to 
$\epsilon = 0.4$ in the equilibrium transition, it can also be 
neglected here. The second cause is the presence of the
nonlinear regimes in the plots of $X_L^Q$ vs $\hat{\chi}_L$ at
low $\hat{\chi}_L$, resulting in non-zero $X_L^Q$-intercepts in the application
of Eq. (\ref{eq:FDR}) to Figs.~\ref{fluct.susc} and \ref{fluct.susc2}. 
Because of this, division of the $X_L^Q$ data by the appropriate 
$T_{\mathrm{eff}}$
values (given in the caption of Fig.~\ref{fluct.susc}) does not quite reproduce
the corresponding $\hat{\chi}_L$ data, 
and (even below $\theta = 0.4$) does not quite result 
in scaling consistent with $\delta_{\mathrm{d}} = 1.75$ with statistical 
signficance. The third and most signficant cause of the variations of
the $\mathcal{G}^X_+$ data is the `doubly linear' behavior observed in 
Fig.~\ref{fluct.susc2} for $\theta > 0.4$, which prevents identification 
of a unique $T_{\mathrm{eff}}$ in this range.

The assumption that $X_L^Q$ can be used as a proxy for $\hat{\chi}_L$ is thus
fairly well justified close to the critical period, where $T_{\mathrm{eff}}$
varies over a limited range and the more complicated effects observed at
$\theta > 0.4$ are not relevant. This is supported by Fig.~\ref{gvsy1.h2},
where the data points cluster closely around the line corresponding to
power-law scaling with exponent $-1.73 \approx -\gamma_{\mathrm{e}}$. 
However, because of the first two causes just described, there are small 
systematic variations in the $\mathcal{G}^X_+$ data which prevent a statistically
significant fit to a pure power-law relationship as a function of $y_1$.

In order to clarify the relationship of these scaling results to those in 
previous work, we also plot in Fig.~\ref{gvsy1.h0.xabsq} data of 
$\mathcal{G}^{|X|}_+(y_1,y_2) \equiv X_L^{|Q|} L^{-\gamma / \nu} 
\equiv \left( \langle Q^2\rangle - \langle \left|Q \right|\rangle^2\right) 
L^{-\gamma / \nu}$ vs $y_1$ at $y_2 = 0$, again using the equilibrium values
$\gamma_{\mathrm{e}}$ and $\nu_{\mathrm{e}}$ to calculate $\mathcal{G}^{|X|}_+$.
This can be directly compared to Fig.~11(d) in Ref. \cite{kn:korniss01},
in which the quantity we call $X_L^{|Q|}$ was called $X_L^Q$. Attempted fits
to all five data points and to the last four data points of 
$\mathcal{G}^{|X|}_+$ in Fig.~\ref{gvsy1.h0} produce nominal scaling 
exponents $-1.60 \pm 0.02$ and $-1.69 \pm 0.02$ (both with $q < 10^{-15}$).
The agreement in Fig.~11(d) of Ref. \cite{kn:korniss01} of the line with slope
$-7/4$ with the data for $\theta > \theta_c$ must therefore be viewed as 
qualitative. The method used in Ref. \cite{kn:korniss01} to numerically 
estimate $\gamma_{\mathrm{d}}$, however, which involves finite-size scaling at
the critical period, is fully consistent with the results of this paper, since
at each period with $\theta < 0.4$ we have found that the FDR in Eq.
(\ref{eq:FDR}) holds to a very good approximation.

\begin{figure}[ht]
\begin{center}
\includegraphics[width=4.5in,angle=270]{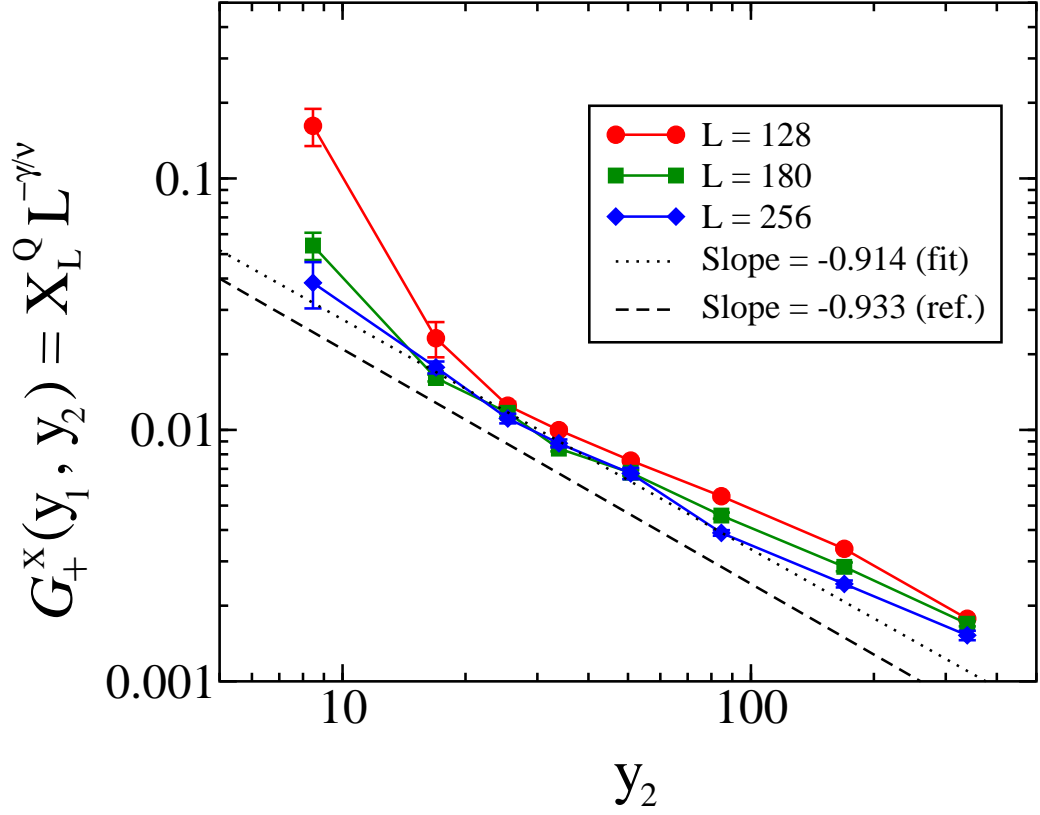} 
\end{center}
\caption{(Color online.) 
Log-log plot of $\mathcal{G}^X_+(y_1,y_2)$ vs 
$y_2 = (H_b/J) L^{\beta \delta / \nu}$ for lattice sizes 
$L = 128,$ 180, and 256, at the critical period $P_c$, where $y_1 = 0$. 
The best-fit line to the $L=256$ data in the range 
$8.46 < y_2 < 84.6$, shown as a dotted line in the plot, corresponds to a scaling exponent 
$(1-\delta_{\mathrm{d}})/\delta_{\mathrm{d}} = -0.914 \pm 0.029$. A reference line (dashed), 
corresponding to a scaling exponent 
$(1-\delta_{\mathrm{e}})/\delta_{\mathrm{e}} = -14/15 \approx -0.933$, is also shown.
}
\label{g+.vs.y2}
\end{figure}

Finally, in Fig.~\ref{g+.vs.y2}, we plot $\mathcal{G}^X_+$ vs $y_2$ at $P=P_c$, 
to study its scaling in 
the regime $y_1 \ll y_2$. As just noted, the use of $X_L^Q$ as a proxy for
$\hat{\chi}_L$ is well justified at $P = P_c$ by our results. 
Power-law scaling is perhaps suggested in the range $20 < y_2 < 50$ for
$L=180$, and it is clearly obeyed from $y_2 = 8.42$ to $84.2$ for $L = 256$. 
The scaling exponent was determined as 
$(1-\delta_{\mathrm{d}})/\delta_{\mathrm{d}} = -0.914 \pm 0.030$, which is 
consistent with the corresponding equilibrium value 
$(1-\delta_{\mathrm{e}})/\delta_{\mathrm{e}} = -14/15 \approx -0.933$. 

\section{Conclusions and outlook}
\label{s:conclusion}

In this article, we have continued the computational study of the dynamic
phase transition (DPT) in the 
two-dimensional kinetic Ising model exposed to a periodically oscillating
field, which was begun in Refs. \cite{kn:sides98,kn:sides99,kn:korniss01}. We have 
established two distinct but related results about the field 
conjugate to the dynamic order parameter. First, we have identified the
period-averaged magnetic field, or `bias field', $H_b$ as an important
component of the full conjugate field. This claim is supported by
numerical evidence that the dynamic order parameter and its susceptibility 
follow critical scaling with respect to $H_b$.
In particular, the scaling exponent $\delta_{\rm d}$ of the conjugate
field was determined for the first time, and found by 
finite-size scaling analysis of large-scale kinetic Monte Carlo simulations
to be equal to the
critical-isotherm exponent for the equilibrium Ising transition,
$\delta_{\rm e} = 15$.
Furthermore, in agreement with previous results \cite{kn:korniss01}, the dynamic
scaling exponents $\gamma_{\rm d}$, $\beta_{\rm d}$, and $\nu_{\rm d}$
were also found to equal their equilibrium Ising counterparts,
$\gamma_{\rm e} = 7/8$, $\beta_{\rm e} = 1/8$, and $\nu_{\rm d} = 1$.

These results further strengthen previous numerical \cite{kn:sides98,kn:sides99,kn:korniss01} and
analytical \cite{kn:fujisaka01,kn:grinstein85,kn:bassler94} claims that the DPT in a periodically driven
two-dimensional kinetic Ising model belongs to the universality class of the 
equilibrium two-dimensional Ising model. However, with respect to the direct 
applicability of the symmetry arguments of Refs.~\cite{kn:grinstein85,kn:bassler94}, 
we caution the reader that what is claimed in the present paper 
(as well as in Ref.~\cite{kn:fujisaka01}) is only equivalence of the 
{\it phase transitions\/} in the driven kinetic Ising model and the equilibrium 
Ising model. {\it Outside\/} the critical region, it is neither clear how closely 
$P - P_c$ and $H_b$ play the roles of $T - T_c$ and the ordinary magnetic 
field, respectively, nor how closely the 
dynamic order parameter, $Q$, corresponds to the 
average equilibrium magnetization. From our discussion of the FDR in 
Sec.~\ref{s:fluct.relation}, it appears likely that one or more of these relations 
break down outside the critical region.  Much theoretical work remains to be done in this area.

The second main result of this article is that a fluctuation-dissipation 
relation (FDR), that 
is, a proportionality relation between the scaled fluctuations $X_L^Q \equiv L^2 \left(
\langle Q^2 \rangle - \langle Q \rangle ^2 \right)$ and the susceptibility
$\hat{\chi}_L$ with a slope we have called $T_{\mathrm{eff}}$, holds 
for a range of periods above $P_c$ and for a range of bias fields
around $H_b = 0$. We stress again that we have found the FDR of Eq.~(\ref{eq:FDR}) 
to hold {\it only in the critical
region} in this nonequilibrium system, in contrast to the equilibrium FDR of 
Eq.~(\ref{eq:FDR0})
which follows directly from the partition function, and which thus holds 
everywhere. We note that, for the parameters used in our computation at least, 
the critical region 
in which the nonequilibrium FDR holds ($P < 190$~MCSS) is somewhat smaller than 
the critical region in which power-law scaling is obeyed ($P < 350$~MCSS).
In previous work, when the conjugate field had not been
identified, the scaled fluctuations $X_L^Q$ were used as a proxy for the
(then unknown) quantity $\hat{\chi}_L$. The evidence for the FDR presented here
shows this assumption to be fully justified at the
critical period (see Fig. \ref{g+.vs.y2}), and to be
a very good approximation -- nearly as 
good as the
use of the scaled fluctuations as a proxy for the 
susceptibility in the equilibrium Ising model -- 
in the critical region where the FDR holds.

There are at least three further computational projects suggested by the progress 
reported here. The first is to investigate whether the field $H_b$ functions
as the conjugate field, with scaling exponents consistent with the equilibrium
Ising transition for periods $P < P_c$, below the critical period. In the
equilibrium system, the study of critical scaling in nonzero field for
$T < T_c$ is complicated by the long time correlations and strong finite-size effects which accompany the bimodal distributions of magnetization below
$T_c$. Similar effects would complicate the 
investigation of scaling with respect to $H_b$ in the DPT for $P < P_c$, but 
the advanced techniques \cite{kn:berg92,kn:berg91} which make
the equilibrium simulations tractable do not extend obviously to the nonequilibrium case.
The second computational project suggested is to determine the nature of the
full conjugate field $H_c$. The third project would be to study the FDR at different
values of the temperature, $T$, and the amplitude, $H_0$, of the driving field.
Finally, we remark that it would be very desirable to extend the
current understanding of the theory of nonequilibrium steady states to include
the conjugate field $H_b$, the FDR found in the critical region, and the
scaling of $H_b$.

\section*{Acknowledgments}

Research at Florida State and Mississippi State Universities was supported by
NSF Grant No.~DMR-0444051, and at Clarkson University by NSF Grant 
No.~DMR-0509104. This research also used resources of 
the National Center for Computational Sciences at Oak Ridge National Laboratory, which is 
supported by the Office of Science of the U.S. Department of Energy under Contract No. 
DE-AC05-00OR22725.

\clearpage 

\begin{thebibliography}{45}
\expandafter\ifx\csname natexlab\endcsname\relax\def\natexlab#1{#1}\fi
\expandafter\ifx\csname bibnamefont\endcsname\relax
  \def\bibnamefont#1{#1}\fi
\expandafter\ifx\csname bibfnamefont\endcsname\relax
  \def\bibfnamefont#1{#1}\fi
\expandafter\ifx\csname url\endcsname\relax
  \def\url#1{\texttt{#1}}\fi
\expandafter\ifx\csname urlprefix\endcsname\relax\def\urlprefix{URL }\fi
\providecommand*{\bibinfo}[2]{#2}
\providecommand*{\eprint}[1]{#1}
\providecommand*{\url}[1]{#1}
\begingroup\makeatletter
 \@temptokena{%
  \expandafter\ifx\csname citenamefont\endcsname\relax
   \DeclareRobustCommand\citenamefont{\@firstofone}%
   \global\let\citenamefont\citenamefont
   \global\expandafter\let\csname citenamefont \expandafter\endcsname\csname
  citenamefont \endcsname
  \fi
 }\if@filesw\immediate\write\@auxout{\the\@temptokena}\fi
\expandafter\endgroup\the\@temptokena

\bibitem[{\citenamefont{Tom\'e and de~Oliveira}(1990)}]{kn:tome90}
\bibinfo{author}{\bibfnamefont{T.}~\bibnamefont{Tom\'e}} \bibnamefont{and}
  \bibinfo{author}{\bibfnamefont{M.~J.} \bibnamefont{de~Oliveira}},
  \bibinfo{journal}{Phys. Rev. A} \textbf{\bibinfo{volume}{41}},
  \bibinfo{pages}{4251} (\bibinfo{year}{1990}).

\bibitem[{\citenamefont{Mendes and Lage}(1991)}]{kn:mendes91}
\bibinfo{author}{\bibfnamefont{J.~F.~F.} \bibnamefont{Mendes}}
  \bibnamefont{and} \bibinfo{author}{\bibfnamefont{E.~J.~S.}
  \bibnamefont{Lage}}, \bibinfo{journal}{J. Stat. Phys.}
  \textbf{\bibinfo{volume}{64}}, \bibinfo{pages}{653} (\bibinfo{year}{1991}).

\bibitem[{\citenamefont{Zimmer}(1993)}]{kn:zimmer93}
\bibinfo{author}{\bibfnamefont{M.~F.} \bibnamefont{Zimmer}},
  \bibinfo{journal}{Phys. Rev. E} \textbf{\bibinfo{volume}{47}},
  \bibinfo{pages}{3950} (\bibinfo{year}{1993}).

\bibitem[{\citenamefont{Acharyya and Chakrabarti}(1995)}]{kn:acharyya95}
\bibinfo{author}{\bibfnamefont{M.}~\bibnamefont{Acharyya}} \bibnamefont{and}
  \bibinfo{author}{\bibfnamefont{B.~K.} \bibnamefont{Chakrabarti}},
  \bibinfo{journal}{Phys. Rev. B} \textbf{\bibinfo{volume}{52}},
  \bibinfo{pages}{6550} (\bibinfo{year}{1995}).

\bibitem[{\citenamefont{Acharyya}(1997{\natexlab{a}})}]{kn:acharyya97b}
\bibinfo{author}{\bibfnamefont{M.}~\bibnamefont{Acharyya}},
  \bibinfo{journal}{Phys. Rev. E} \textbf{\bibinfo{volume}{56}},
  \bibinfo{pages}{2407} (\bibinfo{year}{1997}{\natexlab{a}}).

\bibitem[{\citenamefont{Acharyya}(1998)}]{kn:acharyya98}
\bibinfo{author}{\bibfnamefont{M.}~\bibnamefont{Acharyya}},
  \bibinfo{journal}{Phys. Rev. E} \textbf{\bibinfo{volume}{58}},
  \bibinfo{pages}{179} (\bibinfo{year}{1998}).

\bibitem[{\citenamefont{Lo and Pelcovits}(1990)}]{kn:lo90}
\bibinfo{author}{\bibfnamefont{W.~S.} \bibnamefont{Lo}} \bibnamefont{and}
  \bibinfo{author}{\bibfnamefont{R.~A.} \bibnamefont{Pelcovits}},
  \bibinfo{journal}{Phys. Rev. A} \textbf{\bibinfo{volume}{42}},
  \bibinfo{pages}{7471} (\bibinfo{year}{1990}).

\bibitem[{\citenamefont{Acharyya}(1997{\natexlab{b}})}]{kn:acharyya97a}
\bibinfo{author}{\bibfnamefont{M.}~\bibnamefont{Acharyya}},
  \bibinfo{journal}{Phys. Rev. E} \textbf{\bibinfo{volume}{56}},
  \bibinfo{pages}{1234} (\bibinfo{year}{1997}{\natexlab{b}}).

\bibitem[{\citenamefont{Sides} \emph{et~al.}(1998)\citenamefont{Sides, Rikvold,
  and Novotny}}]{kn:sides98}
\bibinfo{author}{\bibfnamefont{S.~W.} \bibnamefont{Sides}},
  \bibinfo{author}{\bibfnamefont{P.~A.} \bibnamefont{Rikvold}},
  \bibnamefont{and} \bibinfo{author}{\bibfnamefont{M.~A.}
  \bibnamefont{Novotny}}, \bibinfo{journal}{Phys. Rev. Lett.}
  \textbf{\bibinfo{volume}{81}}, \bibinfo{pages}{834} (\bibinfo{year}{1998}).

\bibitem[{\citenamefont{Sides} \emph{et~al.}(1999)\citenamefont{Sides, Rikvold,
  and Novotny}}]{kn:sides99}
\bibinfo{author}{\bibfnamefont{S.~W.} \bibnamefont{Sides}},
  \bibinfo{author}{\bibfnamefont{P.~A.} \bibnamefont{Rikvold}},
  \bibnamefont{and} \bibinfo{author}{\bibfnamefont{M.~A.}
  \bibnamefont{Novotny}}, \bibinfo{journal}{Phys. Rev. E}
  \textbf{\bibinfo{volume}{59}}, \bibinfo{pages}{2710} (\bibinfo{year}{1999}).

\bibitem[{\citenamefont{Chakrabarti and Acharyya}(1999)}]{kn:chakrabarti99}
\bibinfo{author}{\bibfnamefont{B.~K.} \bibnamefont{Chakrabarti}}
  \bibnamefont{and} \bibinfo{author}{\bibfnamefont{M.}~\bibnamefont{Acharyya}},
  \bibinfo{journal}{Rev. Mod. Phys.} \textbf{\bibinfo{volume}{71}},
  \bibinfo{pages}{847} (\bibinfo{year}{1999}).

\bibitem[{\citenamefont{Yasui} \emph{et~al.}(2002)\citenamefont{Yasui, Tutu,
  Yamamoto, and Fujisaka}}]{kn:yasui03}
\bibinfo{author}{\bibfnamefont{T.}~\bibnamefont{Yasui}},
  \bibinfo{author}{\bibfnamefont{H.}~\bibnamefont{Tutu}},
  \bibinfo{author}{\bibfnamefont{M.}~\bibnamefont{Yamamoto}}, \bibnamefont{and}
  \bibinfo{author}{\bibfnamefont{H.}~\bibnamefont{Fujisaka}},
  \bibinfo{journal}{Phys. Rev. E} \textbf{\bibinfo{volume}{66}},
  \bibinfo{pages}{036123} (\bibinfo{year}{2002}).

\bibitem[{\citenamefont{Fujiwara} \emph{et~al.}(2004)\citenamefont{Fujiwara,
  Tutu, and Fujisaka}}]{kn:fujiwara04}
\bibinfo{author}{\bibfnamefont{N.}~\bibnamefont{Fujiwara}},
  \bibinfo{author}{\bibfnamefont{H.}~\bibnamefont{Tutu}}, \bibnamefont{and}
  \bibinfo{author}{\bibfnamefont{H.}~\bibnamefont{Fujisaka}},
  \bibinfo{journal}{Phys. Rev. E} \textbf{\bibinfo{volume}{70}},
  \bibinfo{pages}{066132} (\bibinfo{year}{2004}).

\bibitem[{\citenamefont{Fujiwara} \emph{et~al.}(2006)\citenamefont{Fujiwara,
  Tutu, and Fujisaka}}]{kn:fujiwara06}
\bibinfo{author}{\bibfnamefont{N.}~\bibnamefont{Fujiwara}},
  \bibinfo{author}{\bibfnamefont{H.}~\bibnamefont{Tutu}}, \bibnamefont{and}
  \bibinfo{author}{\bibfnamefont{H.}~\bibnamefont{Fujisaka}},
  \bibinfo{journal}{Prog. Theor. Phys. Suppl.} \textbf{\bibinfo{volume}{161}},
  \bibinfo{pages}{181} (\bibinfo{year}{2006}).

\bibitem[{\citenamefont{Shao} \emph{et~al.}(2004)\citenamefont{Shao, Zhong, and
  Lin}}]{kn:shao04}
\bibinfo{author}{\bibfnamefont{Y.~Z.} \bibnamefont{Shao}},
  \bibinfo{author}{\bibfnamefont{W.~R.} \bibnamefont{Zhong}}, \bibnamefont{and}
  \bibinfo{author}{\bibfnamefont{G.~M.} \bibnamefont{Lin}},
  \bibinfo{journal}{Acta Physica Sinica} \textbf{\bibinfo{volume}{53}},
  \bibinfo{pages}{3165} (\bibinfo{year}{2004}).

\bibitem[{\citenamefont{Acharyya}(2003)}]{kn:acharyya03}
\bibinfo{author}{\bibfnamefont{M.}~\bibnamefont{Acharyya}},
  \bibinfo{journal}{Int. J. Mod. Phys. C} \textbf{\bibinfo{volume}{14}},
  \bibinfo{pages}{49} (\bibinfo{year}{2003}).

\bibitem[{\citenamefont{Acharyya}(2004)}]{kn:acharyya04}
\bibinfo{author}{\bibfnamefont{M.}~\bibnamefont{Acharyya}},
  \bibinfo{journal}{Phys. Rev. E} \textbf{\bibinfo{volume}{69}},
  \bibinfo{pages}{027105} (\bibinfo{year}{2004}).

\bibitem[{\citenamefont{Jang and Grimson}(2003)}]{kn:jang01}
\bibinfo{author}{\bibfnamefont{H.}~\bibnamefont{Jang}} \bibnamefont{and}
  \bibinfo{author}{\bibfnamefont{M.~J.} \bibnamefont{Grimson}},
  \bibinfo{journal}{Phys. Rev. E} \textbf{\bibinfo{volume}{63}},
  \bibinfo{pages}{066119} (\bibinfo{year}{2003}).

\bibitem[{\citenamefont{Jang}
  \emph{et~al.}(2003{\natexlab{a}})\citenamefont{Jang, Grimson, and
  Hall}}]{kn:jang03a}
\bibinfo{author}{\bibfnamefont{H.}~\bibnamefont{Jang}},
  \bibinfo{author}{\bibfnamefont{M.~J.} \bibnamefont{Grimson}},
  \bibnamefont{and} \bibinfo{author}{\bibfnamefont{C.~K.} \bibnamefont{Hall}},
  \bibinfo{journal}{Phys. Rev. B} \textbf{\bibinfo{volume}{67}},
  \bibinfo{pages}{094411} (\bibinfo{year}{2003}{\natexlab{a}}).

\bibitem[{\citenamefont{Jang}
  \emph{et~al.}(2003{\natexlab{b}})\citenamefont{Jang, Grimson, and
  Hall}}]{kn:jang03b}
\bibinfo{author}{\bibfnamefont{H.}~\bibnamefont{Jang}},
  \bibinfo{author}{\bibfnamefont{M.~J.} \bibnamefont{Grimson}},
  \bibnamefont{and} \bibinfo{author}{\bibfnamefont{C.~K.} \bibnamefont{Hall}},
  \bibinfo{journal}{Phys. Rev. E} \textbf{\bibinfo{volume}{68}},
  \bibinfo{pages}{046115} (\bibinfo{year}{2003}{\natexlab{b}}).

\bibitem[{\citenamefont{Machado} \emph{et~al.}(2005)\citenamefont{Machado,
  Buend{\'{\i}}a, Rikvold, and Ziff}}]{kn:machado05a}
\bibinfo{author}{\bibfnamefont{E.}~\bibnamefont{Machado}},
  \bibinfo{author}{\bibfnamefont{G.~M.} \bibnamefont{Buend{\'{\i}}a}},
  \bibinfo{author}{\bibfnamefont{P.~A.} \bibnamefont{Rikvold}},
  \bibnamefont{and} \bibinfo{author}{\bibfnamefont{R.~M.} \bibnamefont{Ziff}},
  \bibinfo{journal}{Phys. Rev. E} \textbf{\bibinfo{volume}{71}},
  \bibinfo{pages}{016120} (\bibinfo{year}{2005}).

\bibitem[{\citenamefont{Buend{\'\i}a}
  \emph{et~al.}(2006)\citenamefont{Buend{\'\i}a, Machado, and
  Rikvold}}]{BUEN06}
\bibinfo{author}{\bibfnamefont{G.~M.} \bibnamefont{Buend{\'\i}a}},
  \bibinfo{author}{\bibfnamefont{E.}~\bibnamefont{Machado}}, \bibnamefont{and}
  \bibinfo{author}{\bibfnamefont{P.~A.} \bibnamefont{Rikvold}},
  \bibinfo{journal}{J.\ Mol.\ Struct.: THEOCHEM}
  \textbf{\bibinfo{volume}{769}}, \bibinfo{pages}{189} (\bibinfo{year}{2006}).

\bibitem[{\citenamefont{Korniss} \emph{et~al.}(2001)\citenamefont{Korniss,
  White, Rikvold, and Novotny}}]{kn:korniss01}
\bibinfo{author}{\bibfnamefont{G.}~\bibnamefont{Korniss}},
  \bibinfo{author}{\bibfnamefont{C.~J.} \bibnamefont{White}},
  \bibinfo{author}{\bibfnamefont{P.~A.} \bibnamefont{Rikvold}},
  \bibnamefont{and} \bibinfo{author}{\bibfnamefont{M.~A.}
  \bibnamefont{Novotny}}, \bibinfo{journal}{Phys. Rev. E}
  \textbf{\bibinfo{volume}{63}}, \bibinfo{pages}{016120}
  (\bibinfo{year}{2001}).

\bibitem[{\citenamefont{Korniss} \emph{et~al.}(2002)\citenamefont{Korniss,
  Rikvold, and Novotny}}]{kn:korniss02}
\bibinfo{author}{\bibfnamefont{G.}~\bibnamefont{Korniss}},
  \bibinfo{author}{\bibfnamefont{P.~A.} \bibnamefont{Rikvold}},
  \bibnamefont{and} \bibinfo{author}{\bibfnamefont{M.~A.}
  \bibnamefont{Novotny}}, \bibinfo{journal}{Phys. Rev. E}
  \textbf{\bibinfo{volume}{66}}, \bibinfo{pages}{056127}
  (\bibinfo{year}{2002}).

\bibitem[{\citenamefont{Chatterjee and Chakrabarti}(2003)}]{kn:chatterjee03}
\bibinfo{author}{\bibfnamefont{A.}~\bibnamefont{Chatterjee}} \bibnamefont{and}
  \bibinfo{author}{\bibfnamefont{B.~K.} \bibnamefont{Chakrabarti}},
  \bibinfo{journal}{Phys. Rev. E} \textbf{\bibinfo{volume}{67}},
  \bibinfo{pages}{046113} (\bibinfo{year}{2003}).

\bibitem[{\citenamefont{Chatterjee and Chakrabarti}(2004)}]{kn:chatterjee04}
\bibinfo{author}{\bibfnamefont{A.}~\bibnamefont{Chatterjee}} \bibnamefont{and}
  \bibinfo{author}{\bibfnamefont{B.~K.} \bibnamefont{Chakrabarti}},
  \bibinfo{journal}{Phase Transitions} \textbf{\bibinfo{volume}{77}},
  \bibinfo{pages}{581} (\bibinfo{year}{2004}).

\bibitem[{\citenamefont{Acharyya}(1999)}]{kn:acharyya99}
\bibinfo{author}{\bibfnamefont{M.}~\bibnamefont{Acharyya}},
  \bibinfo{journal}{Phys. Rev. E} \textbf{\bibinfo{volume}{59}},
  \bibinfo{pages}{218} (\bibinfo{year}{1999}).

\bibitem[{\citenamefont{Fujisaka} \emph{et~al.}(2001)\citenamefont{Fujisaka,
  Tutu, and Rikvold}}]{kn:fujisaka01}
\bibinfo{author}{\bibfnamefont{H.}~\bibnamefont{Fujisaka}},
  \bibinfo{author}{\bibfnamefont{H.}~\bibnamefont{Tutu}}, \bibnamefont{and}
  \bibinfo{author}{\bibfnamefont{P.~A.} \bibnamefont{Rikvold}},
  \bibinfo{journal}{Phys. Rev. E} \textbf{\bibinfo{volume}{63}},
  \bibinfo{pages}{036109} (\bibinfo{year}{2001}); 
  \textbf{\bibinfo{volume}{63}}, \bibinfo{pages}{059903}(E) 
  (\bibinfo{year}{2001}).

\bibitem[{\citenamefont{Tutu and Fujiwara}(2004)}]{kn:tutu04}
\bibinfo{author}{\bibfnamefont{H.}~\bibnamefont{Tutu}} \bibnamefont{and}
  \bibinfo{author}{\bibfnamefont{N.}~\bibnamefont{Fujiwara}},
  \bibinfo{journal}{J. Phys. Soc. Jpn.} \textbf{\bibinfo{volume}{73}},
  \bibinfo{pages}{2680} (\bibinfo{year}{2004}).

\bibitem[{\citenamefont{Meilikhov}(2004)}]{kn:meilikhov04}
\bibinfo{author}{\bibfnamefont{E.~Z.} \bibnamefont{Meilikhov}},
  \bibinfo{journal}{JETP Letters} \textbf{\bibinfo{volume}{79}},
  \bibinfo{pages}{620} (\bibinfo{year}{2004}).

\bibitem[{\citenamefont{Dutta}(2004)}]{kn:dutta04}
\bibinfo{author}{\bibfnamefont{S.~B.} \bibnamefont{Dutta}},
  \bibinfo{journal}{Phys. Rev. E} \textbf{\bibinfo{volume}{69}},
  \bibinfo{pages}{066115} (\bibinfo{year}{2004}).

\bibitem[{\citenamefont{Grinstein} \emph{et~al.}(1985)\citenamefont{Grinstein,
  Jayaprakash, and He}}]{kn:grinstein85}
\bibinfo{author}{\bibfnamefont{G.}~\bibnamefont{Grinstein}},
  \bibinfo{author}{\bibfnamefont{C.}~\bibnamefont{Jayaprakash}},
  \bibnamefont{and} \bibinfo{author}{\bibfnamefont{Y.}~\bibnamefont{He}},
  \bibinfo{journal}{Phys. Rev. Lett.} \textbf{\bibinfo{volume}{55}},
  \bibinfo{pages}{2527} (\bibinfo{year}{1985}).

\bibitem[{\citenamefont{Bassler and Schmittmann}(1994)}]{kn:bassler94}
\bibinfo{author}{\bibfnamefont{K.~E.} \bibnamefont{Bassler}} \bibnamefont{and}
  \bibinfo{author}{\bibfnamefont{B.}~\bibnamefont{Schmittmann}},
  \bibinfo{journal}{Phys. Rev. Lett.} \textbf{\bibinfo{volume}{73}},
  \bibinfo{pages}{3343} (\bibinfo{year}{1994}).

\bibitem[{\citenamefont{Robb} \emph{et~al.}(2006)\citenamefont{Robb, Xu,
  Hellwig, McCord, Berger, Novotny, and Rikvold}}]{kn:robb06}
\bibinfo{author}{\bibfnamefont{D.~T.} \bibnamefont{Robb}},
  \bibinfo{author}{\bibfnamefont{Y.~H.} \bibnamefont{Xu}},
  \bibinfo{author}{\bibfnamefont{O.}~\bibnamefont{Hellwig}},
  \bibinfo{author}{\bibfnamefont{J.}~\bibnamefont{McCord}},
  \bibinfo{author}{\bibfnamefont{A.}~\bibnamefont{Berger}},
  \bibinfo{author}{\bibfnamefont{M.~A.} \bibnamefont{Novotny}},
  \bibnamefont{and} \bibinfo{author}{\bibfnamefont{P.~A.}
  \bibnamefont{Rikvold}}, \emph{\bibinfo{title}{Evidence for a dynamic phase
  transition in [Co/Pt]$_3$ magnetic multilayers}} (\bibinfo{year}{2007}),
  \bibinfo{note}{to be submitted to Phys. Rev. B}.

\bibitem[{\citenamefont{Bander and Mills}(1988)}]{kn:mills88}
\bibinfo{author}{\bibfnamefont{M.}~\bibnamefont{Bander}} \bibnamefont{and}
  \bibinfo{author}{\bibfnamefont{D.~L.} \bibnamefont{Mills}},
  \bibinfo{journal}{Phys. Rev. B} \textbf{\bibinfo{volume}{38}},
  \bibinfo{pages}{R12015} (\bibinfo{year}{1988}).

\bibitem[{\citenamefont{Back} \emph{et~al.}(1995)\citenamefont{Back,
  W{\"{u}}rsch, Vaterlaus, Ramsperger, Maier, and Pescia}}]{kn:back95}
\bibinfo{author}{\bibfnamefont{C.}~\bibnamefont{Back}},
  \bibinfo{author}{\bibfnamefont{C.}~\bibnamefont{W{\"{u}}rsch}},
  \bibinfo{author}{\bibfnamefont{A.}~\bibnamefont{Vaterlaus}},
  \bibinfo{author}{\bibfnamefont{U.}~\bibnamefont{Ramsperger}},
  \bibinfo{author}{\bibfnamefont{U.}~\bibnamefont{Maier}}, \bibnamefont{and}
  \bibinfo{author}{\bibfnamefont{D.}~\bibnamefont{Pescia}},
  \bibinfo{journal}{Nature} \textbf{\bibinfo{volume}{378}},
  \bibinfo{pages}{597} (\bibinfo{year}{1995}).

\bibitem[{\citenamefont{Onsager}(1944)}]{ONSA44}
\bibinfo{author}{\bibfnamefont{L.}~\bibnamefont{Onsager}},
  \bibinfo{journal}{Phys.\ Rev.} \textbf{\bibinfo{volume}{65}},
  \bibinfo{pages}{117} (\bibinfo{year}{1944}).

\bibitem[{\citenamefont{Landau and Binder}(2000)}]{kn:landau00}
\bibinfo{author}{\bibfnamefont{D.~P.} \bibnamefont{Landau}} \bibnamefont{and}
  \bibinfo{author}{\bibfnamefont{K.}~\bibnamefont{Binder}},
  \emph{\bibinfo{title}{A Guide to Monte Carlo Simulations in Statistical
  Physics}} (\bibinfo{publisher}{Cambridge University Press, Cambridge, UK},
  \bibinfo{year}{2000}).

\bibitem[{\citenamefont{Binder and Landau}(1984)}]{kn:binder84}
\bibinfo{author}{\bibfnamefont{K.}~\bibnamefont{Binder}} \bibnamefont{and}
  \bibinfo{author}{\bibfnamefont{D.~P.}~\bibnamefont{Landau}},
  \bibinfo{journal}{Phys. Rev. B} \textbf{\bibinfo{volume}{30}},
  \bibinfo{pages}{1477} (\bibinfo{year}{1984}).

\bibitem[{\citenamefont{Privman and Fisher}(1984)}]{kn:privman84}
\bibinfo{author}{\bibfnamefont{V.}~\bibnamefont{Privman}} \bibnamefont{and}
  \bibinfo{author}{\bibfnamefont{M.~E.} \bibnamefont{Fisher}},
  \bibinfo{journal}{Phys. Rev. B} \textbf{\bibinfo{volume}{30}},
  \bibinfo{pages}{322} (\bibinfo{year}{1984}).

\bibitem[{\citenamefont{Privman}(1990)}]{kn:privman90}
\bibinfo{editor}{\bibfnamefont{V.}~\bibnamefont{Privman}, in 
  \bibfnamefont{V.}~\bibnamefont{Privman}}, ed.,
  \emph{\bibinfo{title}{Finite Size Scaling and Numerical Simulation of
  Statistical Systems}} (\bibinfo{publisher}{World Scientific},
  \bibinfo{address}{Singapore}, \bibinfo{year}{1990}), pp.
  \bibinfo{pages}{4--98}.

\bibitem{scaling.param.endnote}
Until the theory of the DPT, e.g. in Ref. \cite{kn:fujisaka01}, is extended to include $H_c$, it is unclear
in what form it should enter the dimensionless scaling parameter, i.e. as $H_c / J$, $H_c/k_B T$, or otherwise.
Here we have chosen the form $H_c/J$.
Since our numerical data is all taken at a single 
temperature, $T = 0.8T_c = 1.8152 J$, a change in this form would 
at most multiply values of $y_2$ by a constant factor, and would 
not affect the scaling relations reported in this paper.

\bibitem[{\citenamefont{Hayashi and Sasa}(2004)}]{HAYA04}
\bibinfo{author}{\bibfnamefont{K.}~\bibnamefont{Hayashi}} \bibnamefont{and}
  \bibinfo{author}{\bibfnamefont{S.-I.} \bibnamefont{Sasa}},
  \bibinfo{journal}{Phys.\ Rev. E} \textbf{\bibinfo{volume}{69}},
  \bibinfo{pages}{066119} (\bibinfo{year}{2004}).

\bibitem[{\citenamefont{Hayashi and Sasa}(2005)}]{HAYA05}
\bibinfo{author}{\bibfnamefont{K.}~\bibnamefont{Hayashi}} \bibnamefont{and}
  \bibinfo{author}{\bibfnamefont{S.-I.} \bibnamefont{Sasa}},
  \bibinfo{journal}{Phys.\ Rev.\ E} \textbf{\bibinfo{volume}{71}},
  \bibinfo{pages}{046143} (\bibinfo{year}{2005}).

\bibitem{eff.temp.endnote}
To avoid misunderstandings, note that it is certainly possible to 
define other sensible effective temperatures for an Ising-like model, so that
one should be clear about the definition being used in a given context. For example,
the nonequilibrium FDR proposed in Refs.~\cite{GODR00,SAST03} connects
{\it two-time} response and correlation functions, and applies to critical ageing
in a two-parameter family of Ising-like models. This family includes many models without
detailed balance (e.g., the noisy voter and majority models), and one model --
the equilibrium Ising model with Glauber dynamics -- with detailed balance. In contrast,
our Eq.~(\ref{eq:FDR}) concerns {\it single-time} response and correlations, and the kinetic Ising
model driven by an oscillating field is not included in the two-parameter family of models
considered in Refs.~\cite{GODR00,SAST03}.


\bibitem[{\citenamefont{Berg and Neuhaus}(1992)}]{kn:berg92}
\bibinfo{author}{\bibfnamefont{B.~A.} \bibnamefont{Berg}} \bibnamefont{and}
  \bibinfo{author}{\bibfnamefont{T.}~\bibnamefont{Neuhaus}},
  \bibinfo{journal}{Phys. Rev. Lett.} \textbf{\bibinfo{volume}{68}},
  \bibinfo{pages}{9} (\bibinfo{year}{1992}).

\bibitem[{\citenamefont{Berg and Neuhaus}(1991)}]{kn:berg91}
\bibinfo{author}{\bibfnamefont{B.~A.} \bibnamefont{Berg}} \bibnamefont{and}
  \bibinfo{author}{\bibfnamefont{T.}~\bibnamefont{Neuhaus}},
  \bibinfo{journal}{Phys. Lett. B} \textbf{\bibinfo{volume}{267}},
  \bibinfo{pages}{249} (\bibinfo{year}{1991}).

\bibitem[{\citenamefont{Godr{\`e}che and Luck}(2000)}]{GODR00}
\bibinfo{author}{\bibfnamefont{C.}~\bibnamefont{Godr{\`e}che}} \bibnamefont{and}
  \bibinfo{author}{\bibfnamefont{J.~M.}~\bibnamefont{Luck}},
  \bibinfo{journal}{J. Phys. A: Math. Gen.} \textbf{\bibinfo{volume}{33}},
  \bibinfo{pages}{9141} (\bibinfo{year}{2000}).

\bibitem[{\citenamefont{Sastre} \emph{et~al.}(2003)\citenamefont{Sastre, 
Dornic, and Chat{\'e}}}]{SAST03}
\bibinfo{author}{\bibfnamefont{F.}~\bibnamefont{Sastre}},
\bibinfo{author}{\bibfnamefont{I.}~\bibnamefont{Dornic}},
 \bibnamefont{and} \bibinfo{author}{\bibfnamefont{H.}~\bibnamefont{Chat{\'e}}},
  \bibinfo{journal}{Phys. Rev. Lett.} \textbf{\bibinfo{volume}{91}},
  \bibinfo{pages}{267205} (\bibinfo{year}{2003}).


\end{thebibliography}
\end{document}